\documentstyle[pra,aps]{revtex}
\begin{document}
\def\question#1{{{\marginpar{\tiny \sc #1}}}}
\draft
\title{The MSW Effect in Quantum Field Theory}
\author{Christian Y. Cardall}
\address{ Department of Physics \& Astronomy, 
	State University of New York at Stony Brook, 
	Stony Brook, NY 11794-3800
\thanks{Electronic mail: {\tt Christian.Cardall@sunysb.edu}}}
\author{Daniel J. H. Chung}
\address{Randall Physics Laboratory, 
	University of Michigan, Ann Arbor, MI 48109-1120
\thanks{Electronic mail: {\tt djchung@landau.physics.lsa.umich.edu}}}
\date{April 1999}
\maketitle

\begin{abstract}
We show in detail the general relationship
between the Schr\"{o}dinger equation approach to calculating the MSW
effect and the quantum field theoretical S-matrix approach.  
We show the precise form a generic neutrino propagator 
must have to allow a physically meaningful
``oscillation probability'' to be decoupled from neutrino production
fluxes and detection cross-sections,
and
explicitly list the 
conditions---not realized in cases of
current experimental interest---in 
which the field theory approach would be useful.
\end{abstract}

\pacs{14.60.Pq, 26.65.+t, 97.60.Bw, 13.15.+g}

\def\nua{\nu_\alpha}
\def\nub{\nu_\beta}
\def\nuanub{\nu_ \alpha \rightarrow \nu_\beta}
\def\kx{\sum_i k_i}
\def\py{\sum_j p_j}
\def\mat{{\cal M}}
\def\be{\begin{equation}}
\def\ee{\end{equation}}
\def\eqr#1{{Eq.\ (\ref{#1})}}
\def\Xs{{\bf x}_S}
\def\Yd{{\bf y}_D}
\def\eqr#1{{Eq.\ (\ref{#1})}}
\def\eld{\sum_l (-1)^{d_l} E_{\underline {\vp}_l}}
\def\els{\sum_l (-1)^{s_l} E_{\underline {\vk}_l}}
\def\kl{\sum_l (-1)^{s_l} {\underline k_l}}
\def\pl{\sum_l (-1)^{d_l} {\underline p_l}}
\def\pln{\sum_l (-1)^{d_l} p_l}
\def\kln{\sum_l (-1)^{s_l}  k_l}
\def\puln{\sum_l (-1)^{d_l} {\underline p_l}}
\def\kuln{\sum_l (-1)^{s_l} {\underline k_l}}
\def\question#1{{{\marginpar{\tiny \sc #1}}}}
\section{Introduction}
\def\nua{\nu_\alpha}
\def\nub{\nu_\beta}
\def\nuanub{\nu_ \alpha \rightarrow \nu_\beta}
\def\vp{{\bf p}}
\def\vl{{\bf l}}
\def\vk{{\bf k}}
\def\vq{{\bf q}}
\def\vx{{\bf x}}
\def\vy{{\bf y}}
\def\vz{{\bf z}}
\def\vs{{\bf s}}
\def\vr{{\bf r}}
\def\vu{{\bf u}}
\def\bp#1{{\underline p_{#1} }}
\def\bk#1{{\underline k_{#1} }}
\def\enu{E_{\vq}}
\def\gmu{\gamma^\mu}
\def\dmu{\partial_\mu}
\def\lhat{\hat{\bf L}}

Recent results from the Super-Kamiokande experiment 
\cite{sk} have confirmed, 
with high statistics, the reality of the solar 
\cite{solrev} and atmospheric \cite{atmrev} 
neutrino anomalies. In the case of the atmospheric neutrino anomaly,
the Super-Kamiokande data can be interpreted as providing evidence
that the anomaly is due to neutrino flavor mixing \cite{skatmint}, and
``long baseline'' accelerator and reactor neutrino experiments
may be able to
confirm this conclusion \cite{longbl}. Data from
Super-Kamiokande and the Sudbury Neutrino Observatory (SNO) 
eventually should allow tests
to determine if the solar neutrino anomaly is also due to neutrino
oscillations \cite{solsig}. These developments are bringing 
to a critical test the possibility suggested by earlier 
observations of solar and atmospheric neutrinos---and by a
persistent signal in an accelerator neutrino experiment, 
LSND \cite{lsnd}---that neutrino flavor mixing may provide one of the 
first experimental windows on physics beyond the Standard Model.

In a generic neutrino oscillation experiment 
the event rate for the detection of $\nu_\beta$
from a flux of $\nu_\alpha$ from a point source at distance $L$ 
takes a form like the following:
\begin{equation}
d\Gamma_{\alpha\beta} = \int dE_{\vq}  \left(d\Gamma_{\alpha,\nua} \over
	L^2\, d\Omega_{\vq} \, dE_{\vq} \right) \left(P_{\nuanub}\right)
	\left(d\sigma_{\nub,\beta}\right), \label{rate}
\end{equation}
where the direction of the neutrino momentum $\vq$ points
from the source to the detector.
The first factor in the integrand represents the
flux of neutrinos of energy $E_{\vq}$ from a process involving a charged lepton
of flavor $\alpha$, and the third factor is the cross section
for neutrino detection via a process involving a charged lepton
of flavor $\beta$. These factors are computed by the standard techniques
of quantum field theory (QFT),
with the approximation of massless neutrinos. 

The middle factor---the 
so called ``oscillation probability''---is typically computed 
with a quantum mechanical model of the oscillation process. In this 
model the neutrino state inhabits a Hilbert space whose dimension
is equal to the number of neutrino flavors.
The Hilbert space is spanned by a mass basis and a flavor basis. 
These bases are connected 
by a mixing matrix, taken to be the same as that which connects 
neutrino ``flavor fields'' to neutrino ``mass eigenstate fields'' in
a field theory Lagrangian. The Hamiltonian is simply the particle energy,
and the phenomenon of neutrino flavor oscillations arises because the
Hamiltonian is not diagonal in the flavor basis. It was pointed out
by Wolfenstein \cite{wolf} that the parameters of flavor oscillations 
are altered in the presence of matter because of the effective mass induced by 
neutrino forward scattering off the background. The effective mass,
which contributes to the Hamiltonian in this quantum mechanical model,
can be computed by employing the famous formula relating the
index of refraction to the forward scattering amplitude, where this amplitude
is computed from QFT using standard interactions.
  Since the effective mass due to the background is
diagonal in the flavor basis, in the limit of slowly varying background
density the total Hamiltonian is diagonal in a new basis, the
``instantaneous mass'' basis. 
Mikheyev and Smirnov \cite{ms} subsequently noted that a level crossing
of the ``instantaneous neutrino mass eigenstates in matter'' occurs
in a background of monotonically varying density. The resulting
``MSW effect'' constitutes a new mechanism of flavor transformation 
that allows a parameter space of solution to, for example, the solar
neutrino problem, that is very different from that provided 
by vacuum neutrino oscillations.

The means just described of computing an experimental event rate for
neutrino flavor transformations has the virtue of simplicity.  However,
being somewhat schizophrenic in its amalgamation of quantum field
theoretical and quantum mechanical methods, it is not surprising that
studies have appeared in which the neutrino
production/oscillation/detection is examined as a single process in
the context of QFT, with the neutrinos being virtual
particles \cite{rich,grimusstock,camp,kier,cohere,ioapilaf}. 
These studies identify the conditions for which the
amplitude for this overall process factorizes.
However, 
by not showing the complete relationship between
this amplitude and the neutrino production flux and
detection cross section, these studies lack a firm
justification (other than recognition from the usual quantum
mechanical picture) for calling a particular factor
an ``oscillation amplitude.''\footnote{A partial connection
is made in Ref. \cite{grimusstock}, where it is shown, 
for example, how the
$1/L^2$ flux factor and on shell momentum space neutrino 
spinors arise from the vacuum propagator. However, 
the overall event rate they arrive at by ``heuristic consideration,'' 
to use their words, contains a normalization constant. The
relationship of this normalization constant to the coordinate space 
external particle spinors they employ is left unspecified.}  
In addition, previous works 
employing this ``scattering approach'' to flavor mixing
have
only
considered vacuum flavor oscillations; it is our purpose here to
consider the MSW effect in the context of QFT.
Our approach
is complementary to the work of Ref.\ \cite{sireraperez} where this is
derived from QFT within the context of relativistic Wigner functions.

In 
this scattering approach to neutrino
oscillations, the derivation of the usual Schr\"{o}dinger equations for
the MSW effect in the presence of a spatially varying background is a
rather trivial consequence of the virtual neutrinos going on shell,
since in that case the multiparticle nature of QFT becomes irrelevant.
This property was noted in Ref.\ \cite{grimusstock} in the context of
vacuum oscillations.  If this property is invalid, then the usual
quantum mechanical treatment must be modified and the quantum field
theoretical treatment becomes useful.  Hence, perhaps the most
important point of this paper is that the usual quantum mechanical
treatment should be modified suitably if one of the following
conditions holds: the on-shell neutrino momentum is nonrelativistic,
the neutrino production or detection vertices are non-chiral, the wave
packets of the 
external production/detection particles
are sensitive to momentum variations of the order of inverse
source-detector distance, or the neutrino 
effective mass splittings (determined
by the effective potential including the background matter
contributions) are large compared to the spread in the momenta of the
external particles.

We begin in Sec. II with a discussion of general neutrino
oscillations.  While this ground has been 
partially explored previously, 
the
discussion will serve to clarify some of the physics of the
oscillation process and identify the extent to which the neutrino
propagator determines the probability of neutrino oscillations.
In particular, we do not assume the vacuum propagator at the
outset; we find the
precise form a generic propagator must take
to allow a physically meaningful
``oscillation probability'' to be decoupled from neutrino production
fluxes and detection cross-sections, and 
pinpoint the component that gives rise to the oscillation 
amplitude.
In Sec. III, we discuss the effective Lagrangian
approximation assumed in our framework;
as specific examples we discuss $e^\pm$ and neutrino backgrounds,
including an outline of how to obtain a self-consistent neutrino
background in, for example, the supernova environment.
Having identified the Green's
function (or propagator) as that which determines the portion of the
neutrino oscillation probability that is independent of the
production/detection mechanisms, in Sec. IV we study the Green's
function of an effective theory of neutrinos in a static, uniform
background,
finding a rich pole structure. 
Great simplification occurs in the relativistic limit, 
and we recover the same oscillation
amplitude obtained with the usual quantum mechanical model.
In
Sec. V, we study the Green's function in a nonuniform background
potential. Under appropriate conditions we recover the usual
Schr\"{o}dinger-type equation for the oscillation amplitude. 
Sec. VI contains concluding remarks.  An
Appendix contains a quick and easy derivation in QFT
of fully normalized neutrino oscillation event rates in the form of
Eq. (\ref{rate}), which is justified by the more complete treatment in
Sec. II.

\section{GENERAL $\nu$ OSCILLATIONS}
\label{sec:general}
For the sake of completeness and to establish the setting of our
calculation, we give a 
 general overview
of the neutrino oscillation
calculation in field theory,  
going beyond previous works 
\cite{rich,grimusstock,camp,kier,cohere,ioapilaf} in a couple of ways.
 While earlier studies demonstrated the factorization
of the amplitude under suitable conditions---enabling identification of
the factor called the ``oscillation amplitude'' in the standard
picture---they do not go all the way to a fully normalized expression like
that of Eq. (\ref{rate}),
leaving one without an unambiguous, physically meaningful reason to
name this factor an ``oscillation amplitude.''
 In addition, previous works exploring the conditions under which the
neutrino propagator determines the oscillation probability, 
independent of the details of
neutrino production and detection, have only considered the vacuum
propagator. As a prelude to studying the MSW effect,
we seek to elucidate the form a generic neutrino propagator
must
 have
to make it possible to disentangle the
neutrino oscillation probability from the production/detection
mechanisms.

In this section
we 
consider generalities without committing to
a particular model (Lagrangian), establishing the connection between
a physically clear definition
of oscillation probability [given by
\eqr{rate}] and the neutrino propagator.
 For a simple choice of external particle wave packets, 
we will give a fairly general expression
for the neutrino oscillation event rate including a ``calculated''
normalization. 
For illustration, in the Appendix 
we give an explicit calculation leading to a fully
normalized event rate like Eq. (\ref{rate}) in the context of a
semirealistic model Lagrangian.

In field theory and in physical situations, we distinguish a given
flavor of neutrinos by their interactions
  with charged leptons.  
If we calculate the
probability amplitude for the process involving 
flavor $\alpha$ at the source of the neutrinos and 
flavor $\beta$ at the detector of the neutrinos, we can calculate the
probability of neutrino oscillations of $\alpha \rightarrow \beta$.
The flavor of each interaction can be distinguished by measurable, on
shell, external particles
  (i.e., the charged leptons).  
Hence, calculating neutrino oscillations
is equivalent to calculating a scattering event where a neutrino
propagator connects two flavor distinguishing vertices with external
particles coming from them.

In calculating scattering quantities such as cross sections, we
usually calculate the plane wave scattering $S$-matrix
\begin{equation}
S(\{ k_i \},\{ p_j \}  )-1 \equiv (2 \pi)^4 \delta^{4}
\left(\pln+\kln\right) i \mat,
\label{eq:smatrix}
\end{equation}
where $\mat$ is the usual invariant amplitude calculated with Feynman
diagrams in momentum space and $s_l=d_l=1$ for incoming particles and
$0$ for outgoing particles (the grouping of the momenta will be
explained shortly).  This is considered to be a good approximation in
the case of calculating usual collider event rates since there the
events of interest occur within a single volume element before the
final state particles are detected, and the corrections arising from
localization of the interactions usually are not important to the
detection rate.  However, because neutrino oscillations involve
quantum interference effects over macroscopic distances which separate
the production point and the detection point of the neutrinos, we must
take more care to account for the localization of the interaction
points to calculate the leading order observable quantity.

This localization of the vertices necessarily requires
that the incident and final states be spatially localized 
wave packets instead of plane wave states.
(As we shall see, analyses which appear to use only plane wave
external states while restricting spatial integrations in an
ad hoc manner \cite{camp,ioapilaf} 
actually have complicated wave packets buried
beneath the surface.)
Hence, the probability
amplitude\footnote{Our conventions for the metric, gamma matrices,
 and normalizations are the same as
Ref. \cite{peskin}. The wave packet normalization is $\int d^3{\vp}\;
(2\pi)^{-3} \left|\psi(\vp)\right|^2~=~1$. 
A plane wave packet $\psi(\vp,\vp')=
{(2\pi)^3\over\sqrt{V}}\delta^3(\vp-\vp')$, where
$V$ is a volume factor, follows this normalization
convention provided $\left[\delta^3(\vp-\vp')\right]^2$
is interpreted as ${V\over(2\pi)^3}\delta^3(\vp-\vp')$.}
 is a superposition of~\eqr{eq:smatrix},
\begin{equation}
{\cal A} = \int \prod_j^{1 + F_D} [dp_j]\
 \psi_{D j}(p_j, \bp{j})
\prod_i^{I_S+F_S}  [dk_i]\
\psi_{S i}(k_i, \bk{i})\ 
\left[S(\{ k_m \}, \{ p_m \} )-1\right],
\label{eq:amplitude}
\end{equation}
 where $[dp_j]=d^3 {\vp}_j/\left[ (2 \pi)^3 \sqrt{2 E_{{\vp}_j}}\right]$, 
$\{ k_m \}$ 
are
the external momenta of the vertex at the production region centered
about $\Xs$, and 
$\{ p_m \}$
are the external momenta of the detection
region centered about $\Yd$.  Here the set of parameters 
$\bp{m}$
and
$\bk{m}$
characterize the peak of the wave packets' distribution of
momenta.\footnote{The 0th component of these ``parameters'' is taken
to be on mass shell since we will choose the wave packets such that
they behave like plane waves for large values of the spatial
components of these ``parameters.''}  We have also fixed the number of
external particles to be $I_S$ incoming and $F_S$ outgoing particles
at the source vertex, and 1 incoming and $F_D$ outgoing external
particles at the
detector vertex. 

Now, let us see how this is related to the usual quantum mechanical
treatment.
The procedure for calculating $P_{\nuanub}$ in \eqr{rate} in
a quantum mechanical model was described in Sec. I. 
 In field theory, $P_{\nuanub}$ defined by
\eqr{rate} can be calculated by comparing \eqr{rate} with the event
rate derived from \eqr{eq:amplitude}.
We associate ${\cal A}_S$ with
the amplitude for 
$\{I_S\} \rightarrow \{F_S\} + \stackrel{(-)}{\nu_\alpha}$ 
at the source
and ${\cal A}_D$ with the amplitude for 
$D + \stackrel{(-)}{\nu_\beta} \rightarrow \{F_D\}$ 
at
the detector (we specialize to one detector particle $D$), 
and choose plane wave packets for the source's final state 
(anti)neutrinos
and the detector's initial state
(anti)neutrinos:
\begin{eqnarray}
{\cal A}_S & =& {1\over\sqrt{2\enu}\sqrt{V}}\int 
\prod_i^{I_S+F_S}  [dk_i]\
\psi_{S i}(k_i, \bk{i})\ 
\left[S_S(\{ k_m \}, q )-1\right], \nonumber\\
{\cal A}_D &=& {1\over\sqrt{2\enu}\sqrt{V}}
\int \prod_j^{1 + F_D} [dp_j]\
 \psi_{D j}(p_j, \bp{j})
\left[S_D(\{ p_m \},q )-1\right],
\label{detamp}
\end{eqnarray}
where $q=(\enu,\hat{\bf L}\enu)$ is the neutrino
momentum, and $\hat{\bf L}=(\Yd - \Xs)/|\Yd -\Xs|$
points from the source to the detector.
In ${\cal A}_S$ and ${\cal A}_D$ we have implicitly assumed
that the neutrinos are massless, because massive flavor eigenstates
cannot be asymptotic states. 
Standard
kinematics then yields the relationship
\begin{equation}
\frac{\left| \cal A \right|^2}{T}=\int \frac{d\enu\, \enu^2}{(2 \pi)^3
L^2 v_{\nu D}} \frac{|{\cal A}_S|^2 V}{T_S} P_{
\stackrel{(-)}{\nu_\alpha}\rightarrow\stackrel{(-)}{\nu_\beta}} 
\frac{|{\cal A}_D|^2 V}{T_D},
\label{eq:relationship}
\end{equation}
where $V$ is the usual total volume factor associated with
the phase space and normalization of plane wave packets; $T$, $T_S$,
and $T_D$ are the usual time factors associated with stationary
wave packets;
$v_{\nu D}$ is the M$\o$ller speed (associated with the flux) between the
detector particle and the neutrinos;
and $L \equiv |\Yd -\Xs|$.  

As just indicated, we make the simplifying assumption of stationary
wave packets.  The main simplifying utility of this energy
conservation approximation is to get rid of the neutrino momentum
integral.\footnote{We refer the reader to Ref.\ \cite{cohere} and
references therein for related discussions regarding coherence.}
We encode our assumption of stationarity by defining spatially smeared
functions
\begin{eqnarray}
g_S(\vx,\{\bk{i}\},q)\; e^{i \kuln\cdot x} & = & \int
\prod_j^{I_S+F_S} [ dk_j]\; \psi_{S j}(k_j, \bk{j})\; e^{i \kln \cdot x}\;
i {\cal M}_S\left(\{k_i \}, q \right),  \nonumber \\
g_D(\vy,\{\bp{i}\},q)\; e^{i \puln\cdot y} & = & \int
\prod_j^{1+F_D} [ dp_j]\; \psi_{D j}(p_j, \bp{j})\; e^{i \pln \cdot y}\;
i {\cal M}_D(\{p_i \}, q). 
\label{eq:stationary}
\end{eqnarray}
We note that ${\cal M}_S$ and ${\cal M}_D$ have the form
(assuming V-A lepton currents)
\begin{eqnarray}
{\cal M}_S\left(\{k_i \}, q \right)&=& \bar u^-(q)P_R M_1(\{k_i\}),
\ \ \ {\cal M}_D(\{p_i \}, q)= M_2(\{p_i\})P_L u^-(q)\ \ \ 
(\nu\ \rm{osc.}), \nonumber\\
{\cal M}_S\left(\{k_i \}, q \right)&=& M_2(\{k_i\})P_L v^+(q), 
\ \ \ {\cal M}_D(\{p_i \}, q)= \bar v^+(q)P_R M_1(\{p_i\}) \ \ \ 
(\bar\nu\ \rm{osc.}), \label{msmd}
\end{eqnarray}
where $M_1$ and $M_2$ are respectively column and row vectors
in spinor space, and $P_L$ and $P_R$ are the left- and right-handed
chiral projection operators.
Following the conventions of \cite{peskin}, we represent
the spinors $u$ and $v$ as
\begin{equation}
u^s(q) =\pmatrix{\sqrt{q\cdot\sigma}\;\xi^s \cr 
\sqrt{q\cdot\bar\sigma}\;\xi^s},
\ \ \ \ v^s(q)=\pmatrix{\sqrt{q\cdot\sigma}\;\eta^s \cr 
-\sqrt{q\cdot\bar\sigma}\;\eta^s},
\end{equation}
where $\sigma^\mu=(1, \mbox{\boldmath$\sigma$} )$,
$\bar\sigma^\mu=(1,-\mbox{\boldmath$\sigma$})$, and
$\mbox{\boldmath$\sigma$}$ is the three-vector of Pauli matrices. 
Since these are spinors for massless particles, the spin
index $s$ is associated with the spin component along the
momentum axis; specifically, we have $\xi^- = \eta^+$, 
$\xi^+ = -\eta^-$, with $\mbox{\boldmath $\sigma$}\cdot\hat\vq\;
\xi^\pm(\hat\vq) =\pm \xi^\pm(\hat\vq)$.  
Putting Eqs. (\ref{detamp},\ref{eq:stationary}) 
into \eqr{eq:relationship} yields
\begin{equation}
(2 \pi)\, \delta\left(\els+\eld\right)\, 
P_{\stackrel{(-)}{\nu_\alpha}\rightarrow\stackrel{(-)}{\nu_\beta}}  
= 16 \pi^2 L^2 v_{\nu D}\,
\frac{\left| \cal{A} 
\right|^2}{T}\, \left| \tilde{g}_S(-\vq, \{ \bk{i} \}) \right|^{-2}
\left| \tilde{g}_D(\vq, \{ \bp{i} \}) \right|^{-2},
\label{eq:almost}
\end{equation}
where we have defined
\begin{equation}
\tilde{g}_D(\vq, \{\bp{i}\})\equiv \int d^3\vy\, g_D(\vy, \{ \bp{i}\},q )
\, e^{i \left(\vq-\sum_l(-1)^{d_l}{\underline{\bf p}_l}\right) \cdot \vy},
\label{gtilde}
\end{equation}
and similarly for $\tilde{g}_S$. Stationarity constrains 
$|\vq|\equiv\enu=-\els$.

Let us turn our attention to $\cal A$.  
Given that we have a neutrino
propagator $G$ in our amplitude, and assuming V-A lepton currents,
we can write 
\begin{equation}
S(\{ k_i \}, \{ p_j \}  )-1 = \int d^4y\,e^{i \pln \cdot y}
	\int d^4x\,e^{i \kln \cdot x}\, i\int {d^4s\over (2\pi)^4}
	e^{ \mp i s\cdot(y-x)} M_2\, P_L\, G(s)\, P_R\, M_1,
\label{eq:factorit}
\end{equation}
where 
$M_1$ and $M_2$ are the same as in Eqs. (\ref{msmd}), and
$s$ is the off-shell propagator momentum. 
The upper (lower) sign of $\mp$ in the
exponential is for neutrino (antineutrino) oscillations;
this arises from choosing $x$ ($y$) to always
correspond to the source (detector). 
That is, for neutrino oscillations of flavor $\alpha$
to flavor $\beta$, the Green's function is 
$iG^{\beta\alpha}(y,x)=\langle T\{\nu^\beta(y)\bar\nu^\alpha(x)\}\rangle_0
=i\int{d^4s\over(2\pi)^4}\,e^{-is\cdot(y-x)}G^{\beta\alpha}(s)$
(with $T\{\}$ and $\langle \rangle_0$ denoting a time-ordered product
and vacuum expectation value respectively),
while for antineutrino oscillations $\alpha\rightarrow \beta$,
the labeling is $iG^{\alpha\beta}(x,y)$. 
 
Insert the identity
\begin{equation}
{q^\mu \xi^\nu \{\gamma_\mu,\gamma_\nu\} \over 2q\cdot \xi}=1
\label{eq:ident1}
\end{equation}
on both sides of the Green's function in equation \eqr{eq:factorit},
where as before $q=(\enu,\,\hat{\bf L}\enu)$, and we define
$\xi= (\enu,\, \hat{\bf L}\sqrt{(\enu)^2 - m^2})$ 
in which the parameter 
$m^2 < \enu^2$, though its precise value is unimportant in this context.
We note that since $q$ is null, $q\cdot \gamma = \sum_s u^s(q)\bar u^s(q)$
(or $q\cdot \gamma = \sum_s v^s(q)\bar v^s(q)$, if one wishes to consider
antineutrino oscillations).
From the explicit form of $u$ and $v$ it is easy
to see that
\begin{eqnarray}
P_L\, (q\cdot\gamma)(\xi\cdot\gamma)&=&P_L\, u^-(q)\bar u^-(q)\,
 (\xi\cdot\gamma),
\ \ \ \ P_L\, (\xi\cdot\gamma)(q\cdot\gamma)=
	(\xi\cdot\gamma)\,P_R\, u^+(q)\bar u^+(q),\nonumber\\
(\xi\cdot\gamma)(q\cdot\gamma)\, P_R &=& (\xi\cdot\gamma)\,
	u^-(q)\bar u^-(q)\,P_R,
\ \ \ \ (q\cdot\gamma)(\xi\cdot\gamma)\, P_R=
	u^+(q)\bar u^+(q)\, P_L\, (\xi\cdot\gamma). \label{spinors}
\end{eqnarray}
[The same relations hold for $u^\pm(q)$ replaced by $v^\mp(q)$.]
Soon we will show the form that the Green's function 
must have, after localization by the source and detector, in order
that the term with 
$u^-$ (or $v^+$)
on both sides of $G$ be
the only one to contribute. If more than one spin
contributes, we will not recover the usual quantum mechanical
treatment without spins taken into account.  In that case, working
with the full scattering picture of \eqr{eq:amplitude} is useful.
 Keeping only the 
term with the relevant spinor on either side of $G$,
one can
show that Eq. (\ref{eq:amplitude}) becomes
\begin{equation}
{\cal A}= -\int d^4y \; g_D(\vy,\{\bp{i}\},q)\, e^{i \puln\cdot y}
	\int d^4x\; g_S(\vx,\{\bk{i}\},q)\, e^{i \kuln\cdot x}\,
	i\int {d^4s\over (2\pi)^4}
	e^{\mp i s\cdot(y-x)} \bar P\, G(s)\, P, \label{amplitude2} 
\end{equation}
where 
$P=\gamma^0 u^-(q)/(2\enu) = \gamma^0 v^+(q)/(2\enu) $, 
$\bar P=P^\dagger \gamma^0$,
and $g_S$ and $g_D$ are given
by Eq. (\ref{eq:stationary}).

In passing, we would like to remark that in
\eqr{amplitude2}, we can always write 
(for neutrino oscillations, for example)
\begin{equation}
g_S(\vx, \{ \bk{i} \},q)\, e^{i \kuln\cdot x} = i \bar{u}^-(q)\, M_1(\{ \bk{i} 
\})\, f_S(x, \{ \bk{i} \} ) 
\end{equation}
(and similarly for $g_D$) where $f_S$ is a scalar function of
$\vx$. Then we obtain the form (again assuming the single-spin
contribution is justified after spatial integration)
\begin{equation}
{\cal A} = \int d^4x\, d^4y\, f_S(x, \{\bk{i} \}) f_D(y, \{ \bp{i}
\} ) \int \frac{d^4s}{(2 \pi)^4} e^{-i s \cdot (y-x) } i {\cal M}(\{
\bk{i} \}, s, \{ \bp{i} \} ),
\label{eq:usual}
\end{equation}
which is a common starting point of analysis in the literature as
in Refs. \cite{camp,ioapilaf}.  However, 
with arbitrarily chosen smearing functions $f$, it is difficult to
assess what actual scattering question the amplitude is an answer to,
because the smearing functions are not the wave 
functions of the in-out particles, but are the wave functions smeared
over the matrix elements.  As a consequence, the normalization is
usually ignored in this approach.  We will return to the normalization
later in this section.

Returning to Eq. (\ref{amplitude2}), integration over $x^0$, $y^0$,
and $s^0$ gives an overall energy-conserving delta function and
sets $s^0 = \enu$ 
(for antineutrino oscillations,
$s^0 = -\enu$). We also note that the chiral structure of 
$\bar P\, G P$ (as well as the original matrix element) 
picks out only the $G_{LR}$ block of the neutrino propagator,
where $G_{LR}$ is the nonzero $2\times2$ submatrix left by
$P_L G P_R$. 
In addition, the localization of $g_S$ and $g_D$ around
$\Xs$ and $\Yd$ respectively ``clamps down'' on the coordinate
space Green's function. 
In particular, if the characteristic widths $L_S$ and $L_D$ of
$g_S$ and $g_D$ are much smaller than the source-detector
distance $L=|\vy_D-\vx_S|$, we 
 note that if the ``oscillation
probability'' is to be disentangled from the details of neutrino
production and detection, 
the relevant portion of the Green's function
for oscillations $\alpha\rightarrow\beta$ 
must
take the form
\begin{eqnarray}
G_{LR}^{\beta\alpha}(s^0=\enu,\vy,\vx) &=& \int {d^3{\bf s}\over (2\pi)^3}\,
e^{i\vs \cdot (\vy-\vx)}\,G_{LR}^{\beta\alpha}(s^0=\enu,\vs)\nonumber\\
& \simeq& -\enu \left(1-\mbox{\boldmath $\sigma$}\cdot \hat{\bf L}\right)  
 {e^{i\enu\hat{\bf L}\cdot(\vy-\vx)} \over 4\pi |\vy_D-\vx_S|}
  H^{\beta\alpha}(\enu,\Yd,\Xs)\ \ \ \ \ (\nu\ \rm{osc.}), \nonumber\\
& & \nonumber\\
G_{LR}^{\alpha\beta}(s^0=-\enu,\vx,\vy) 
& \simeq& +\enu \left(1-\mbox{\boldmath $\sigma$}\cdot \hat{\bf L}\right)  
 {e^{i\enu\hat{\bf L}\cdot(\vy-\vx)} \over 4\pi |\vy_D-\vx_S|}
  \bar H^{\alpha\beta}(\enu,\Yd,\Xs)\ \ \ \ \ (\bar\nu\ \rm{osc.}), 
\label{gform}
\end{eqnarray}
where 
$\hat{\bf L} = (\vy_D-\vx_S)/|\vy_D-\vx_S|$ points from the source towards
the detector,
and
the quantities
$H$ and $\bar H$ have
only flavor 
indices.
The
factor $\enu \left(1-\mbox{\boldmath $\sigma$}\cdot \hat{\bf L}\right)$
arises from the kinetic term in the Lagrangian, and
takes this form due to
the relativistic limit. 
Another key
ingredient
is the factor $e^{i\enu|\vy-\vx|}$, which 
is the leading phase factor in the
relativistic limit coming from $e^{ i \vs \cdot (\vy-\vx)}$ evaluated
at the 
poles
of $G_{LR}$.  
In addition,
$1/|\vy-\vx|$ comes from the
asymptotic expansion of the left hand side of \eqr{gform} in the limit
that $|\vy-\vx| \rightarrow \infty$, and it can be considered to be
the monopole term in a multipole expansion.  We will discuss the
validity of the factorization and the asymptotic expansion further
below in momentum space.

Before we talk about momentum space, let us
give an example of \eqr{gform} by considering the vacuum propagator.
In that case, it is straightforward to show that 
(anticipating the relativistic limit)
\begin{eqnarray} 
G_{LR}(s^0,\vx,\vy)
	& =& \left(s^0 + i\mbox{\boldmath $\sigma$}\cdot
\mbox{\boldmath$\nabla$}\right)\left[M^{-1}
	G_{RR}(s^0,\vx,\vy)\right], \label{glrgrr}\\
\left[M^{-1} G_{RR}(s^0,\vx,\vy)\right]^{\alpha\beta}&=&   
- {e^{i|s^0||\vx-\vy|} \over 4\pi |\vx-\vy|}
 \sum_j U_{\alpha j}U_{\beta j}^* \exp\left(-i{m_j^2|\vx-\vy|\over
	2|s^0|}\right), \label{grr} 
\end{eqnarray}
where 
$G_{RR}$ is the nonzero $2\times2$ submatrix left by
$P_R G P_R$, $M$ is the mass matrix appearing in the Lagrangian,
and the $m_j$ are the mass eigenvalues. 
In Eq. (\ref{grr}) we have made the flavor indices explicit; 
the relationship between the
flavor fields and mass eigenstate fields is $\nu_\alpha = \sum_i
U_{\alpha i} \psi_i$, where the $U_{\alpha i}$ are elements of a
unitary matrix. 
For $|s^0|L\gg 1$,
\begin{equation}
G_{LR}^{\alpha\beta}(s^0,\vx,\vy)\simeq
- {e^{i|s^0||\vx-\vy|} \over 4\pi |\vx-\vy|}
\left[s^0 -|s^0|\mbox{\boldmath $\sigma$}\cdot
\hat\vr(\vx,\vy)\right]
\sum_j U_{\alpha j}U_{\beta j}^* \exp\left(-i{m_j^2|\vx-\vy|\over
	2|s^0|}\right), \label{glrfinal}
\end{equation}
where $\hat\vr(\vx,\vy)=(\vx-\vy)/|\vx-\vy|$.
To apply Eq.\ (\ref{glrfinal}) to neutrino oscillations one takes
$s^0\rightarrow\enu$ and
$\vx,\vy \rightarrow \vy,\vx$. For antineutrino oscillations,
$s^0\rightarrow -\enu$ and
$\vx,\vy\rightarrow\vx,\vy$.
After making these substitutions 
we will be integrating Eq. (\ref{glrfinal}) over
localization functions 
of characteristic widths
$L_S$ and $L_D$ centered on $\vx=\vx_S$
and $\vy=\vy_D$. 
This means that for
$L_S,L_D \ll L$, 
we may replace $\vx$ and $\vy$ by $\vx_S$ and $\vy_D$
everywhere except the phase factors, in which we consider
the first order variation,
\begin{eqnarray}
|\vx-\vy|&\simeq& | \vx_S-\vy_D | + \lhat \cdot \left[  (\vy-\vy_D) 
	- (\vx-\vx_S) \right] \nonumber\\
&= &\lhat \cdot (\vy-\vx).
\label{dapprox}
\end{eqnarray}
We see that
a necessary mathematical condition 
(in addition to the relativistic assumption
and $\enu L\gg 1$)
for
Eqs. (\ref{glrgrr}-\ref{glrfinal})
to reduce to the form of Eqs. (\ref{gform}) 
is $m_j^2\, L_{S,D} /(2\enu) \ll 1$ [which also implies
the more familiar $(m_j^2-m_i^2)\, L_{S,D} /(2\enu) \ll 1$]. 
The physical basis of these conditions
can also be inferred from Eqs. (\ref{glrgrr}-\ref{glrfinal}).
$\enu L\gg 1$ allows the propagating neutrino to become
an on-shell relativistic particle, and also allows appreciable
oscillation phase to build up over the source-detector distance. 
$m_j^2\, L_{S,D} /(2\enu) \ll 1$  requires
that no appreciable oscillation phase build up on length
scales comparable to the width of the external particle wave packets.
The necessity of these conditions for disentanglement of the
flavor oscillations from the details of neutrino production and
detection is evident.

The origin of these conditions can 
also
be
understood in momentum space.  First, rewrite \eqr{amplitude2} as
\begin{eqnarray}
{\cal A} & = & - (2 \pi) \delta\left(\els +
\eld\right) i \int \frac{d^3 s}{(2 \pi)^3} 
e^{ \pm i {\bf s} \cdot (\Yd - \Xs)} e^{ -i \left(\sum_l (-1)^{s_l}
{\underline {\bf k}_l} \cdot \Xs + \sum_l (-1)^{d_l} {\underline {\bf
p}_l} \cdot \Yd \right )}\nonumber\\ & &\times
\tilde{h}_{D}({\bf s}, \{ \bp{i}  \}, q)
\tilde{h}_{S}(- {\bf s}, \{ \bk{i} \}, q) \bar{P}\; 
G\left(s^0 = \mp\els ,{\bf s} \right) P
\label{eq:a}
\end{eqnarray}
where the functions $\tilde{h}_{S}$ and $\tilde{h}_{D}$ can be
approximated to have no ${\bf x}_S$ or ${\bf y}_D$ dependence.
This can easily be seen to be exactly true for the ideal case of isotropic
smearing functions, e.g. $g_D(\vy, \{ \bp{i}\},q )= h_D(|\vy-\vy_D|)$;
Eq. (\ref{gtilde}) then yields $\tilde{g}_D = e^{i\vu\cdot\vy_D} 
\tilde h_D(|\vu|)$, where  $\vu 
\equiv \vs-\sum_l(-1)^{d_l}{\underline{\bf p}_l}$.
Because the propagator will in general
have poles corresponding to the mass of the physical neutrino states,
the dominant contribution to the integral in the asymptotic limit $ L
E_{\vq} \rightarrow \infty$ will be from a term that contains the
integrand of \eqr{eq:a}
as a factor evaluated at the poles and stationary phase points
(critical points).
  For the vacuum, the constant potential,
and the adiabatically spatially varying background potential cases,
one can asymptotically expand the integral
\eqr{eq:a} in the limit $LE_{\vq} \rightarrow \infty$
(similarly as 
in Ref. \cite{grimusstock}) to find that 
to leading approximation, the term $\tilde{h}_{D}(\vs, \{ \bp{i} \},
q) \tilde{h}_{S}(-\vs, \{ \bk{i} \},q)$ 
can
be moved outside of the integral with the replacement $\vs \rightarrow
\vs_*$ where $\vs_*$ corresponds to one of the critical points.  By
factoring out $\tilde{h}$, we have implicitly assumed that
$\tilde{h}_{D}(\vs, \{ \bp{i} \},q) \tilde{h}_{S}(-\vs, \{ \bk{i}
\},q)$ is not sensitive to the splittings in the critical points
(otherwise different pole momenta ${\bf s}$ of $G_{LR}$ will cause
$\tilde{h}_{D}(\vs, \{ \bp{i} \},q) \tilde{h}_{S}(-\vs, \{ \bk{i}
\},q)$ to have different values, preventing factorization).  This
means that the wave packets must be flat in momentum space at least
within the range of pole momentum splitting.  
Also, if this is not the case, one of the poles will not contribute
(because the amplitude of the wave packet has fallen off with respect
to the amplitude at the other pole), and no neutrino oscillations will
occur  (or more accurately, the neutrino
oscillations will be greatly suppressed relative to the background).
We will refer to this flatness of the wave packet as insensitivity to
$\vs_*$ splitting.

Also note that because of the presence of the exponential in Eq. (18),
this leading term in the asymptotic expansion            
will not be a good approximation unless the inverse
``momentum scale height'' (i.e. logarithmic derivative) of
$\tilde{h}_{D}\,\tilde{h}_{S}$ near the poles is much less than L.
Hence, factoring out the wave packet dependence which is crucial for
the validity of the usual quantum mechanical treatment requires the
wave packet factor $\tilde{h}_{D}\,
\tilde{h}_{S}$ to be insensitive under $1/L$
momenta variations as well as the $\vs_{*}$ splitting variations.

While localization is clearly necessary for the observation of oscillations,  
the source and detector localization scales $L_S,L_D$ implied by Eq. 
(\ref{eq:stationary}) cannot be smaller than the Compton wavelength
of the lightest external particles. In the case that all the external particles
connected to a given vertex are nonrelativistic, this 
gives rise to a constraint on the masses of these external particles.
To see this, 
consider the ideal case mentioned above 
in which $g_D(\vy, \{ \bp{i}\},q )\approx h_D(|\vy-\vy_D|)$.
Then  $\tilde{g}_D = e^{i\vu\cdot\vy_D} 
\tilde h_D(|\vu|)$, where $\tilde h_D(|\vu|)$ is damped for $|\vu| 
\equiv |\vs_*-\sum_l(-1)^{d_l}{\underline{\bf p}_l}|$ larger than $1/L_D$.
 Hence the $\vs_*$ splitting insensitivity
condition can be written as
\begin{eqnarray}
\left| \sum_l (-1)^{s_l} {\underline \vk_{l}} + \vs_* \right| & \ll &
\frac{1}{L_S}    <  M_{LS} \nonumber \\
\left| \sum_l (-1)^{d_l} {\underline \vp_{l}} - \vs_* \right| & \ll &
\frac{1}{L_D}  <  M_{LD}.
\label{eq:osccond1}
\end{eqnarray}
where we have denoted the lightest external particle masses to be
$M_{LS}$ and $M_{LD}$ for source and detector, respectively.
The critical momentum will 
be  $\vs_*
\approx \hat{\bf L} \sqrt{(\els)^2 - \tilde{m}_{j}^2}$ where $\tilde{m}_j$
is the effective pole mass of the particle.  Hence, if $\tilde{m}_{j}
\ll \left|\els\right|$ and the external particles are nonrelativistic, then
\eqr{eq:osccond1} can be satisfied only if about equal mass of
external particles enter and leave the source/detector vertices.  If
any of the
the external particles 
connected to a given vertex
are sufficiently relativistic, this severe
constraint does not arise.

We
now show that Green's function must take the
form found in Eqs. (\ref{gform}) after being spatially ``clamped'' by the
source and detector if the terms projected by the 
spinors 
$u^-$ (or $v^+$)
on both sides of $G$ 
are to
be the only contributions to the
amplitude.   In Sec. \ref{sec:cback}, where we study the
neutrino propagator in a uniform, static medium, we will find it
convenient to identify 
$\lhat$ of Eq. (\ref{gform}) with
the positive third spatial direction.  In that case it is  
straightforward to show,
using Eq. (\ref{spinors}),
that the terms with 
spinors of different spins on either side of $G$
pick out
the off-diagonal 
spinor space
elements of $G_{LR}$, while the terms
with 
the same spins
on both sides pick out the
diagonal 
spinor space
elements.
The matrix $\left(1-\sigma^3\right)$ from Eqs. (\ref{gform}) confirms that
only $G_{LR}^{22}$ is nonzero, and therefore only the term with 
$u^-(q)$ 
[$v^+(q)$] on both sides of $G$ survives for the neutrino
(antineutrino) oscillations.\footnote{Note that with
$\hat{\bf L}$ set to the third spatial direction, both $u^-(q)$ and
$v^+(q)$ have 4-spinor components $(0,\sqrt{2 \enu},  0, 0)$. }

Recalling that $\vq=\hat{\bf L}\,\enu$, and upon inserting Eqs. (\ref{gform})
into Eq. (\ref{amplitude2}) and Eq. (\ref{amplitude2}) into Eq. 
(\ref{eq:almost}),
we finally arrive at the neutrino oscillation probability
\begin{eqnarray}
P_{\nuanub} &=&  
\left|H^{\beta\alpha}(\enu,\Yd,\Xs) \right|^{2},\ \ \ \ (\nu\ \rm{osc.}),
	 \nonumber\\
P_{\bar\nu_\alpha\rightarrow\bar\nu_\beta} &=&  
\left|\bar H^{\alpha\beta}(\enu,\Yd,\Xs) \right|^{2},\ \ \ \ 
	(\bar\nu\ \rm{osc.}),
\label{oscprob}
\end{eqnarray}
where $H$ and $\bar H$ are defined by Eqs. (\ref{gform}),
and we have assumed that the detector particle $D$ is nonrelativistic
such that $v_{\nu D}=1$.
With the cancellation of the source and detector wave packets, one can see
why employing a separate quantum mechanical model to compute
the oscillation probability is possible. [Note that the standard
vacuum oscillation probability is recovered here, 
as is clear from Eqs. (\ref{gform}-\ref{glrfinal})].

We emphasize that
the Green's function here is the full propagator in any given theory and
we have made no severe assumptions about the nature of the
production and detection effective
vertex.\footnote{The most significant 
assumptions leading to our final result were the V-A type of lepton currents;
the stationary approximation in \eqr{eq:stationary}; relativistic
neutrinos, i.e. $(\enu - |\vs_*|)/\enu \ll 1$, where $|\vs_*|$
is the magnitude of a pole in the momentum space propagator; 
sufficiently localized and separated source and detector, i.e.
$(\enu - |\vs_*|) L_{S,D} \ll 1$; and $\enu L \gg 1$.
}  Hence, as
expected, the production/detection independent field theoretical effects
on neutrino oscillations come from the coordinate space Green's function.  
What perhaps is less expected is the fact that unless the
wave packets of the external particles satisfy specific properties,
the transition probability will not just depend on the propagator, but
the entire coherent scattering process which neutrino oscillation really is.
Such tangled wave packet dependence is discussed for example in Ref.\
\cite{ioapilaf}.

Before concluding this section, we would like to note that we can
easily work out the neutrino oscillation detection rate including the
normalization if we assume a particular class of wave packets.
Let us define a box wave packet as a configuration such that the 
superposition integral gives, for each
outgoing particle, for example,\footnote{We
will write the wave packet centered about $\Xs$, since the one
centered about $\Yd$ is analogous.}
\begin{equation}
\int \frac{d^3 k}{(2 \pi)^3 \sqrt{2 E_k}} \psi_{S}(k, {\underline k})
e^{i k \cdot x}\approx
e^{iE_{\underline{\vk}}x^0}
\int \frac{d^3 k}{(2 \pi)^3 \sqrt{2 E_k}} \psi_{S}(k, {\underline k})
e^{-i \vk \cdot \vx}
 = N e^{i {\underline k} \cdot x} B(\vx -\Xs),\label{box}
\end{equation}
where $N$ is a constant independent of $x$ and  
$B(\vz)$
 is a
function which vanishes if 
$\vz$ is outside of a box centered
about the origin with each dimension of length 
$L_S$ and is 1 everywhere else.
The approximation in Eq. (\ref{box}) is valid for
$x^0 \ll (E_{\underline{\vk}} L_S)/(2\pi|\underline{\vk}|)$.
The normalization condition $\int \frac{d^3k}{(2 \pi)^3}
|\psi(k)|^2 =1$ fixes $N$ and implies 
\begin{equation}
\psi_{S}(k)= {1\over\sqrt{V}}
 \left[ 1- {\cal O}\left(\frac{1}{L^2
E_{\underline{\vk}}^2}\right)\right] e^{i(\vk- {\underline\vk}) \cdot \Xs}
D_S(\vk-{\underline\vk}),
\label{eq:normal}
\end{equation}
where
\begin{equation}
D_S({\bf v}) = 8\, \frac{ \sin(v_x L_S/2) \sin(v_y L_S/2) \sin(v_z L_S/2)}{v_x
v_y v_z}.
\label{eq:boxtrans}
\end{equation}
Hence, since the box scale must be larger than the Compton wavelength
scale, these external particles will generally have ``plane wave in a
box'' type of normalization 
[up to a $(2 \pi)^3
\delta^{3}(\vk-{\underline\vk})$ type of localization factor
$D_S$].  
This wave packet can be used to calculate the event rate using
\eqr{eq:amplitude} in a standard way.  Since 
$\psi_S$
 will have a width
$2 \pi/L_S$,
 smearing of any function that is proportional to momenta
whose magnitude at the peak of the distribution is of the order 
$2 \pi/L_S$,
 (or less) will deviate significantly from the Dirac $\delta$
smearing of that function.  Fortunately, because $M_1$ and $M_2$ do
not depend on such small momenta, we can write the amplitude in the
form of \eqr{eq:usual} with
\begin{equation}
f_{S}(x, \{\bk{i} \})= \prod_j^{I_S+F_S} \left[ \frac{e^{-i(-1)^{s_j}
{\underline k_j} \cdot x}}{\sqrt{2 E_{{\underline\vk}_j} V_S}}
B(\vx-\Xs) \right] 
\label{eq:separate}
\end{equation}
(and similarly for 
$f_{D}$) which is what one would use to calculate
scattering of particles confined to a box interacting with particles
that can propagate outside of the box.  
 Explicitly, the transition rate per source and detector particle is
given by 
\begin{eqnarray}
d \Gamma & = &(2 \pi)\, \delta\left(\els + \eld\right) 
\prod_j^{I_S+F_S}\frac{1}{2
E_{{\underline{\bf k}_{j}}} V_S } 
\prod_i^{F_D+1}\frac{1}{ 2 E_{{\underline{\bf p}_{i}}} V_D}
\prod_b^{F_S}  \frac{d^3 {\underline{\bf k}}_{b} V_S}{(2 \pi)^3} 
\prod_a^{F_D}  \frac{d^3 {\underline{\bf p}}_{a} V_D}{(2 \pi)^3} 
  \nonumber \\
& &\times\left| \int \frac{d^3 \vs}{(2 \pi)^3}  e^{\pm i \vs \cdot
(\Yd -\Xs)} D_S\left(\sum_l (-1)^{s_l} {\underline{\bf k}}_l \pm \vs\right)
    D_D\left(\sum_l (-1)^{d_l} {\underline{\bf p}_l} \mp
\vs\right) i {\cal M} \right|^2, 
\end{eqnarray}  
where $D_S$ is defined by \eqr{eq:boxtrans},
and $D_D$ is similarly defined.
In this case, one can also use the usual heuristic box
quantization formalism to calculate the event rates including the
normalization.  For pedagogical purposes we carry out this simple
exercise explicitly using 
a fermion field toy model in
the Appendix.

To summarize this 
section, we have shown to what extent the
neutrino ``oscillation probability'' is determined by the
production/detection wave packet-independent propagator of the field
theory.  If wave packets for the production and
detection events described in \eqr{rate} satisfy suitable localization
properties and the effective mass splitting of the neutrinos is not
large compared to the momentum width of the wave packets,
the neutrino propagator determines the probability of transition as
defined by \eqr{rate}.  This factoring of the wave packets out of the
transition amplitude is crucial to recover the usual quantum
mechanical picture of neutrino oscillations.  Furthermore, we see how
the multiparticle nature of the field theory becomes irrelevant as the
poles of the propagator are the only states to contribute in this
limit.  For this factorization to be possible, the source-detector
separation $L$ must be large enough such that the wave packets do not
vary over $1/L$ momentum perturbations about the pole momentum, and
the pole momenta splitting must be small enough such that the wave
packet amplitudes take on approximately the same value for the various
pole momenta [as discussed between \eqr{eq:a} and \eqr{eq:osccond1}].
Furthermore, since the usual quantum mechanical treatment neglects the
spin of the neutrinos, only one spin projection of the Green function
must contribute to the amplitude to recover the usual treatment.  We
have seen in this section that the relativistic limit of the on shell
neutrinos and the chiral nature of the interactions ensure this.  Now
that we see that wave packet dependence can be factored out (as is
implicit in the usual simple quantum mechanical treatment), we shall
concentrate on the wave packet-independent field theoretic calculation
of the MSW effect, which is encoded in the propagator within a
background medium.

\section{Effective Lagrangian}
In this section, we briefly explain the effective potential employed
in our calculation of the MSW effect.  Focusing on the physics well
below the electroweak scale, we write the usual electroweak effective
Hamiltonian density as (see for example \cite{kuorev1})
\begin{equation}
{\cal H}_I= \frac{G_F}{\sqrt{2}} \left( J_c^{\mu} J_{c \mu}^{\dagger} +
J_N^{\mu} J_{N \mu} \right), \label{ewham}
\end{equation}
where $J_c^{\mu}$ is the charged current and $J_N^{\mu}$ is the
neutral current.  Take for example the contribution to the
neutrino-electron interaction of the form
\begin{equation}
{\cal H}_I = \frac{4 G_F}{\sqrt{2}}  \bar{\nu}_e \gamma^\mu P_L
\nu_e \bar{e} \gamma_\mu P_L e,
\label{eq:enuint}
\end{equation}
which will be dominant for the MSW effect.  We can distinguish two
different types of scattering: forward scattering, for which the
background particles do not change their momenta; and non-forward
scattering.  In calculating our transition rate, we will not account
for 
non-forward scattering contributions
because these can be considered to be 
separate production events. 
With this restriction, in expanding the S-matrix perturbatively,
 the main background contribution will come from the
expectation values of \eqr{eq:enuint} taken with respect to the
electron background states.  The scattering amplitude will then receive
contributions proportional to powers of
\begin{equation}
\langle  V_\mu^e  \rangle \equiv 2 \sqrt{2} G_F \langle n |
\bar{e} \gamma_\mu P_L e| n \rangle 
\end{equation}
where $n$ labels some many-body background electron state (not
necessarily translationally invariant).  Note that the right hand side
is proportional to the left handed electron current of state $|n
\rangle$.
In an experimental setting, we are really interested in ensemble
averages of the probabilities (not the averages of the $S$
matrix).  However, for
macroscopic numbers of electrons, we expect the main contribution to
come from a set of degenerate states having the same spatial
localization as the 
macroscopic
distribution function.  This
approximation will break down if the density matrix is not sharply
peaked about one set of states giving degenerate contributions to the
scattering amplitude.
 We will assume that such a peaked distribution
exists, and we will merely assign 
macroscopic
currents to the
expectation value of currents that will arise in the scattering
amplitude calculation.  
This means that
 we will replace the interaction Hamiltonian
density of \eqr{eq:enuint} with the effective density
\begin{equation}
{\cal H}_I^{\rm eff} = \frac{4 G_F}{\sqrt{2}}  \bar{\nu}_e \gamma^\mu P_L
\nu_e J_\mu^e \label{nueeff}
\end{equation}
where $J_\mu^e$ is the 
macroscopic
left-handed electron current.
For example, for an unpolarized $e^{\pm}$ background one
would employ---based on consideration of the sum over spin states
of single particle expectation 
values of $\bar e \gamma^\mu P_L e$, for example---the
following expression:
\begin{equation}
J_\mu^e = {1\over 2}\int {d^3\vp\over (2\pi)^3}\left[
	f_{e^-}(\vp)-f_{e^+}(\vp)\right]{p_\mu\over E_{\vp}},
\end{equation} 
where the $f_{e^\pm}(\vp)$ are the usual distribution functions,
including a factor of two for spin degeneracy.
As usual, this procedure neglects higher order
correlations.
Note that for electrons in thermal equilibrium, our prescription, e.g.
\begin{equation}
\langle V_\mu^e \rangle = \sqrt{2}G_F(n_{e^-}-n_{e^+}) \delta_{\mu 0}
\end{equation}
gives the same mass shift as 
the real time thermal field formalism
employed by \cite{notraf}.

As another example, the effective potential due to background neutrinos
is of interest in the envelope of a supernova/nascent neutron star, where
the neutrino flavor composition can affect, for example, 
the explosion mechanism \cite{fuller} or the outcome of possible heavy 
element nucleosynthesis \cite{qf1,qf2}. In addition to the $e^\pm$ background,
we must consider neutrino-neutrino forward scattering arising from
another term in Eq. (\ref{ewham}),
\begin{equation}
{\cal H}_I = {G_F\over \sqrt{2}}\sum_{i,j}\bar{\nu}_i \gamma^\mu P_L
\nu_i \bar{\nu}_j \gamma_\mu P_L \nu_j, \label{nunu}
\end{equation}
where the indices $i,j$ label the mass eigenstate fields.
We work in the mass basis because the external
neutrino background consists of on shell states, a point whose
consequences were emphasized in Ref. \cite{sigl} (see also Ref.
{\cite{qf2}, and references in these).
As seen previously, in the perturbative expansion of the S-matrix we
will have occasion to take a background expectation value of
this interaction (this time with respect to a many-body 
background neutrino state). Two of the neutrino fields will
be paired with fields in the ``production'' and ``detection''
interactions, leaving two other fields whose background
expectation value is taken:
\begin{equation}
\langle {\cal H}_I \rangle = 
{G_F\over \sqrt{2}}\sum_{i,j}2 \bar{\nu}_i \gamma^\mu P_L
\nu_i \langle \bar{\nu}_j \gamma_\mu P_L \nu_j\rangle
	+ {G_F\over \sqrt{2}}\sum_{i,j} 2 \bar{\nu}_i \gamma^\mu P_L
\langle\nu_i \bar{\nu}_j\rangle \gamma_\mu P_L \nu_j. \label{nunuave}
\end{equation}
While the correspondence of the expectation value in the first
term of Eq. (\ref{nunuave}) with a macroscopic current is apparent,
the meaning of the second term is less clear. 

We seek guidance
by considering the expectation values with respect to the single
particle neutrino states $|\vq\,s\,\nu_k\rangle$ of momentum $\vq$,
spin $s$, and mass $m_k$. 
The expectation value in the first term of Eq. (\ref{nunuave})
is
\begin{equation}
\langle\vq\,s\,\nu_k | \bar\nu_j\gamma^\mu P_L \nu_j 
	|\vq\,s\,\nu_k\rangle = {\delta_{jk}\over(2\pi)^3(2\enu)}
	\bar u(\vq\,s\,\nu_k)\gamma^\mu P_L u(\vq\,s\,\nu_k).\label{eval1}
\end{equation}
We consider a relativistic neutrino background, so that to leading
order there are only negative helicity states; then the momentum
space spinors in Eq. (\ref{eval1}) are approximately
\begin{equation}
u(\vq\,s\,\nu_k) \simeq \pmatrix{\sqrt{2\enu}\xi^-(\hat\vq)\cr 0},
	\ \ \ \ \ \xi^-(\hat\vq) = \pmatrix{-\sin(\theta/2)e^{-i\phi}
	\cr \cos(\theta/2)}, \label{spinor}
\end{equation}
where $(\theta,\phi)$ denote the polar and azimuthal angles that
define $\hat\vq$.
The first term in Eq. (\ref{nunuave}) becomes
\begin{equation}
{G_F\over \sqrt{2}}\sum_{i,j}2 \bar{\nu}_i \gamma^\mu P_L
\nu_i \langle \bar{\nu}_j \gamma_\mu P_L \nu_j\rangle =
\sqrt{2} G_F \sum_{i,j} \chi_i^\dagger {\delta_{jk}\over
(2\pi)^3}{q_\mu \bar\sigma^\mu \over \enu}\chi_i,\label{term1}
\end{equation}
where $\chi_i$ denotes the upper two components of $P_L\nu_i$
and $\bar\sigma^\mu = (1,-\mbox{\boldmath $\sigma$})$.
Turning to the second term in Eq. (\ref{nunuave}),
one finds
\begin{equation}
\langle\vq\,s\,\nu_k | \nu_i\bar \nu_j 
	|\vq\,s\,\nu_k\rangle = -{\delta_{ik}\delta_{jk}
	\over(2\pi)^3(2\enu)}
	 u(\vq\,s\,\nu_k)\bar u(\vq\,s\,\nu_k).
\end{equation}
Employing Eq. (\ref{spinor}), the second term in Eq. (\ref{nunuave})
becomes
\begin{equation}
{G_F\over \sqrt{2}}\sum_{i,j}2 \bar{\nu}_i \gamma^\mu P_L
\langle \nu_i \bar{\nu}_j\rangle \gamma_\mu P_L \nu_j =
\sqrt{2} G_F \chi_k^\dagger {\delta_{ik}\delta_{jk}\over
(2\pi)^3}{q_\mu \bar\sigma^\mu \over \enu}\chi_k.\label{term2}
\end{equation}
Noting the similarity between Eqs. (\ref{term1}) and (\ref{term2}),
in the relativistic limit we replace the Hamiltonian density of
Eq. (\ref{nunu}) with the effective density
\begin{equation}
{\cal H}_I^{\rm eff} = \sqrt{2}G_F \sum_{i,j}\bar\nu_i
	J_{\nu_j}^\mu \gamma_\mu
	P_L  \nu_i + \sqrt{2}G_F \sum_i
	\bar\nu_i J_{\nu_i}^\mu\gamma_\mu
	P_L  \nu_i, \label{nunueff}
\end{equation}
where
\begin{equation}
J_{\nu_i}^\mu = \int {d^3\vp\over (2\pi)^3}\left[
	f_{\nu_i}(\vp)-f_{\bar\nu_i}(\vp)\right]{p^\mu\over E_{\vp}}.
\label{jnu}
\end{equation}
The flavor fields $\nu_\alpha$ 
are related to the mass eigenstate fields $\nu_i$
by $\nu_\alpha = \sum_i U_{\alpha i}\nu_i$, where
the $U_{\alpha i}$ are elements of a unitary matrix.
In the limit of vanishing mixing angles $(U_{\alpha i}=
\delta_{\alpha i})$ and a thermal background, the effective
interaction of Eq. (\ref{nunueff}) gives the same mass shifts
as obtained in Ref. \cite{notraf}.
For nontrivial mixing, however,
in terms of the flavor fields Eq. (\ref{nunueff}) becomes
\begin{equation}
{\cal H}_I^{\rm eff} = \sqrt{2}G_F \sum_{\alpha,j}
	\bar\nu_\alpha J_{\nu_j}^\mu\gamma_\mu
	P_L \nu_\alpha + \sqrt{2}G_F \sum_{\alpha,\beta}
	\bar\nu_\alpha\left(U_{\alpha i} J_{\nu_i}^\mu 
	U_{i\beta}\right)\gamma_\mu
	P_L \nu_\beta. \label{nunueff2}
\end{equation}
Considered as a matrix in flavor space, the quantity in parentheses
in the second term of Eq. (\ref{nunueff2}) contains
off-diagonal elements. It is clear from our derivation that 
the presence of these off-diagonal terms
derives from the fact that the background neutrinos are
mass eigenstates, since they must be on shell. This origin
of off-diagonal flavor space terms in the background 
potential due to neutrino-neutrino scattering was  
pointed out in Ref. \cite{sigl}.

In calculations of the effects of neutrino flavor oscillations in
the supernova environment, the supernova core is typically treated
as a stationary source of neutrinos free-streaming from a 
``neutrinosphere,'' with the flux at the neutrinosphere being
taken from large-scale numerical computations. 
Since the neutrinos forming the background also undergo flavor
transformation, self-consistency between the oscillation probability
and the background must be achieved. 
Following the framework of Ref. \cite{sigl}, such a calculation
was carried out in Ref. \cite{qf2} in the quantum mechanical picture
of neutrino oscillations. This involved a rather complicated
procedure involving a 
flavor basis density matrix to describe the
neutrinos above the neutrinosphere.
Casual inspection of the form of the second term in 
Eq. (\ref{nunuave}) would seem to make this kind of approach necessary.
However, having shown that this term can plausibly be written in
terms of a macroscopic mass basis current, 
we see that the self-consistency between neutrino background
and oscillation probability is most easily achieved by working in
the mass eigenstate basis.\footnote{It would seem reasonable
to define neutrino flavor distribution functions in the relativistic
limit. While {\em at the emission point} (i.e. the neutrinosphere)
one could argue that these would be related to the mass basis 
distribution functions by $f_{\nu_i}(\vp) = \sum_\alpha 
|U_{\alpha i}|^2 f_{\nu_\alpha}$, at points above the neutrinosphere
the relation between these sets of distribution functions is
rather complicated, due to the flavor/mass oscillations of 
free-streaming neutrinos in a background. The resulting absence of 
a simple connection
between the macroscopic flavor and mass neutrino currents at
arbitrary position to plug into Eq. (\ref{nunueff2}) makes working in
the mass basis seem much more straightforward.}
Given effective interaction Hamiltonians like
Eqs. (\ref{nueeff}) and (\ref{nunueff}), the oscillation probability can be
computed (in any basis) 
as described in Secs. \ref{sec:cback} and \ref{sec:vback}. 
We write Eq. (\ref{jnu}) as 
\begin{equation}
J^\mu_{\nu_i}(r)= 
\int {E_{\vp}^2\, d E_{\vp}\,  d(\cos\theta) d\phi \over
	(2\pi)^3}\,\left[ f_{\nu_i}(E_{\vp},\cos\theta,r) 
 - f_{\bar\nu_i}(E_{\vp},\cos\theta,r)\right]
{p^\mu\over E_{\vp} },  
\end{equation}
where 
 $p^\mu=(E_{\vp},\,\hat\vp E_{\vp} )$, 
$r$ is the radial position of a point above the neutrinosphere,
and $\theta$ is the
angle between the neutrino momentum and the radial
direction at the point with radial position $r$. 
Since the neutrinos
are free-streaming, the distribution functions at $r$ can be expressed
simply in terms of the ``known''  neutrino
distribution functions at the neutrinosphere, e.g.:
\begin{equation}
f_{\nu_i}(E_{\vp},\cos\theta,r)=\sum_{\nu_j}
	f_{\nu_j}(E_{\vp},\cos\psi,R)\, P_{\nu_j
	\rightarrow \nu_i}(E_{\vp},\cos\theta,r),
\end{equation}
where $R$ is the radius of the neutrinosphere, and $\psi$
is the angle of the neutrino emission with respect to 
the radial direction at the emission point; this angle is 
related to $\theta$ by
 $\cos\psi = \sqrt{1-[(r/R)\sin\theta]^2}$. 
The dependence of the oscillation probabilities
on path length (and the background encountered on
a particular path) are implicit in the $r,\theta$
dependence. 
With an iterative procedure,
self-consistency between the macroscopic
neutrino currents and the oscillation
probabilities should be achieved.

\section{Constant Background}
\label{sec:cback}

Having constructed effective interaction Hamiltonians as
described in the last section, 
we 
employ the neutrino effective Lagrangian
\begin{equation}
{\cal L} = \bar\nu \left[\gmu(i\dmu - V_\mu P_{L}) - M\right] \nu, 
\end{equation}
where $M$ is the mass matrix and $V_\mu$ is a uniform background potential
matrix.
We
make no assumptions about the number of neutrino generations
or the structure of the potential matrix (other than to keep in mind that
it might be singular). The canonical anticommutation relations yield the
equation satisfied by the Green's function $G(x,y)$,
\begin{equation}
\left[\gmu(i\dmu - V_\mu P_{L}) - M\right] G(x,y) = \delta^4(x-y),
\label{green}
\end{equation}
where $i\,G(x,y)\equiv \langle T{\psi(x)\bar\psi(y)}\rangle_0$.
With our convention for the $\gamma$ matrices it is convenient
to define the $2\times 2$ (in spinor space) 
matrices $G_{IJ}$, the ``chiral blocks''
of the Green's function. Specifically, $G_{IJ}$ is the nonzero
$2\times 2$ submatrix of $P_I G P_J$, where $I,J$ can take the values
$L,R$. 

In Sec. II we noted that, with the assumption of V-A interactions,
$G_{LR}$ is the object of interest. 
We also saw in Sec. II that with the assumption of stationarity it
is natural to Fourier transform the time variable while maintaining
interest in the spatial dependence of the Green's function. 
Defining $J = M^{-1} G_{RR}$, from Eq. (\ref{green}) we find
\begin{equation}
G_{LR}(\omega,\vx,\vy) = (\omega + i\mbox{\boldmath $\sigma$} \cdot 
	\mbox{\boldmath$\nabla$}) 
	J(\omega,\vx,\vy), \label{glr}
\end{equation}
where $f(x,y) = \int{d\omega\over 2\pi}\, e^{-i\omega(x^0-y^0)}
 f(\omega, \vx,\vy)$.
In the context of neutrino oscillation experiments we are in interested
in well-separated source and detector positions, so we ignore terms
in $J$ with more than one factor of $|\vx-\vy|$ in the denominator, 
\begin{eqnarray}
J(\omega,\vx,\vy) &=& \int {d^3 p \over (2\pi)^3}\, e^{i\vp \cdot
	(\vx-\vy)} \,J(\omega, \vp) \nonumber \\
&=& \int_0^{\infty} {du \, u^2 \over (2\pi)^3} \int_0^{2\pi} d\phi \,
{1\over i u |\vx-\vy|} \left[ e^{iu|\vx-\vy|} J(\omega,u,\hat\vp=+\hat\vr)
	\right.\nonumber\\
& &\left. - e^{-iu|\vx-\vy|} J(\omega, u,\hat\vp=-\hat\vr) + 
	{\cal O}\left(J\over u|\vx-\vy|\right)\right], \label{j1}
\end{eqnarray}
where we have integrated the $\cos\theta$ integral by parts, and 
defined $u\equiv |\vp|$ and $\hat\vr\equiv (\vx-\vy)/|\vx-\vy|$. 
It is evident that the two leading terms are azimuthally symmetric, and
that their the sum is even in $u$. 
Furthermore, the Feynman boundary conditions should ensure
that the two leading terms give equal contributions. We then have
\begin{equation}
J(\omega,\vx,\vy) \simeq {1\over (2\pi)^2 i |\vx-\vy|} 
	\int_{-\infty}^{\infty} du\, u\, e^{iu|\vx-\vy|} 
   J(\omega,u,\hat\vp=+\hat\vr). \label{jxy}
\end{equation}

From Eq. (\ref{green}), we find that $J$ obeys the momentum
space equation
\begin{equation}
\left[\omega^2 - |\vp|^2 - M^2 - \omega V^0 + \vp \cdot {\bf V}
 + \mbox{\boldmath$\sigma$} 
\cdot \left(V^0\vp - \omega{\bf V} + i {\bf V} \times
	{\bf p}\right)\right] J(\omega,\vp) = 1, \label{jmom}
\end{equation}
or $D(\omega,u,\hat\vp=+\hat\vr) J(\omega,u,\hat\vp=+\hat\vr) = 1$.
Detailed expressions for the spinor space elements of 
$D(\omega,u,\hat\vp=+\hat\vr)$
for general orientation of $\vr$ are not particularly illuminating.
However, it is easy to formally express 
the spinor space elements of $J(\omega,u,\hat\vp=+\hat\vr)$
in terms of the elements of $D$:
\begin{eqnarray}
J^{11}(\omega,u,\hat\vp=+\hat\vr)&=& \left[ D^{11} - D^{12}(D^{22})^{-1}
	D^{21}\right]^{-1}, \label{j11}\\
J^{22}(\omega,u,\hat\vp=+\hat\vr)&=& \left[ D^{22} - D^{21}(D^{11})^{-1}
	D^{12}\right]^{-1}, \label{j22}\\
J^{12}(\omega,u,\hat\vp=+\hat\vr)&=& -(D^{11})^{-1}D^{12}J^{22},\\
J^{21}(\omega,u,\hat\vp=+\hat\vr)&=& -(D^{22})^{-1}D^{21}J^{11}.
\end{eqnarray}
These results are valid without any relativistic limit
assumptions.
Given specific mass and potential matrices, one could solve explicitly
for $J(\omega,u,\hat\vp=+\hat\vr)$. To study neutrino oscillations, 
we then need $J(\omega,\vx,\vy)$, whose behavior is seen from Eq. (\ref{jxy})
to be determined by the
poles of $J(\omega,u,\hat\vp=+\hat\vr)$ with positive imaginary part
(as determined by the Feynman boundary conditions).

A few general comments regarding these poles are in order. Consider for example
$J^{22}$,
which can be expressed 
\begin{equation}
J^{22}(\omega,u,\hat\vp=+\hat\vr) = ({\rm det\ } D^{11})
	\left[({\rm det\ } D^{11}) D^{22} - D^{21}(C^{11})^T
	D^{12}\right]^{-1},
\end{equation}
where $(C^{11})^T$ is the transpose of the matrix of cofactors of
$D^{11}$. Since the diagonal elements of $D^{11}$ and $D^{22}$ are
second order in $u$, $({\rm det\ } D^{11})$ is of order $2n$ in
$u$, where $n$ is the number of neutrino generations; and overall the
denominator of $J^{22}(\omega,u,\hat\vp=+\hat\vr)$ will be a polynomial
of order $4n$ in $u$. This is sensible in terms of a quasiparticle picture
associated with the propagator: Each neutrino field, with two spin states
each for particles and antiparticles, represents four states. For a single
vacuum field the masses of these four states are degenerate; however, 
the presence of a parity
and rotational invariance
violating potential breaks this degeneracy.

Let us examine the simplifications that occur in the relativistic limit
and with source and detector localization.  
In 
the relativistic
limit the poles contributing to the integral in Eq. (\ref{jxy})
take the form
\begin{equation}
u \simeq |\omega| - {\tilde m^2\over 2|\omega|} + i\epsilon,
\end{equation}
with the Feynman boundary conditions imposed by giving the
``masses'' $\tilde m^2$ a small negative imaginary part. (There are also
negative poles, with negative imaginary parts, that do not contribute
to the integral; these factors each become $\simeq 2|\omega|$
when evaluated at the positive poles.) 
Furthermore, following the discussion of Sec. \ref{sec:general}
regarding spatial localization, 
Eq. (\ref{jxy}) takes the
form
\begin{equation}
J(\omega,\vx,\vy) \simeq {2|\omega|\,e^{i \omega \lhat \cdot(\vx-\vy)}
	\over 4\pi |\vx_S-\vy_D|} 
	  \sum_j 
e^{-i{\tilde m_j^2\over2|\omega|}|\vx_S-\vy_D|}
 \left.  \left[\left(u - |\omega| + {\tilde m_j^2\over 2|\omega|}\right)
	J\left(\omega,u,\hat\vp=\frac{\omega}{|\omega|}\lhat\right)\right]
\right|_{u\rightarrow
	|\omega| - {\tilde m_j^2\over 2|\omega|}}, \label{jxy2}
\end{equation}
where the sum is over the poles with positive imaginary parts.
We recall that $\omega$ is fixed 
by energy delta functions, to a positive value
for neutrino oscillations and a negative value 
for antineutrino oscillations.
Because the spatial localization sets 
$\hat\vr=\pm \lhat$ 
(with $\lhat$ taken to be the third spatial 
direction), the matrix $D$ takes the relatively simple spinor space form
\begin{equation}
\left[D\left(\omega,u,\hat\vp=\frac{\omega}{|\omega|} \lhat\right)\right]=
\pmatrix{d -\omega\left(1 - {u\over|\omega|}\right)V^0 + \omega
\left({u\over|\omega|}-1\right)V^3 & 
	-\omega \left(1 + {u\over|\omega|}\right)(V^1 - iV^2) \cr
 -\omega\left(1 - {u\over|\omega|}\right)(V^1 + iV^2) & 
d - \omega\left(1 +
{u\over|\omega|}\right)V^0 + \omega\left( {u\over|\omega|} + 1\right)V^3}. 
\label{dmatrix}
\end{equation}
where $d \equiv \omega^2-u^2-M^2$.

Next we examine the momentum space pole structure of 
$J^{22}$ in the relativistic limit. In 
\eqr{j22},
since the residues of only the positive poles contribute, we can replace $u$ by
$|\omega|$ whenever it multiplies a component of $V^\mu$, committing errors
of only ${\cal O}(\omega V^0 /\omega^2)$ or less with respect to 
other terms present. In that case, $D_{21}$ vanishes and 
$J^{22}\rightarrow (D^{22})^{-1}$
with 
\begin{equation}
D^{22} \rightarrow  \omega^2 - u^2 - M^2 - \frac{\omega}{|\omega|} 2 q \cdot V 
\label{d22}\\
\end{equation}  
where $q =(|\omega|,\lhat |\omega|)$ is the same as the neutrino
momentum defined just below \eqr{detamp}.
We note that the denominator of $J^{22}$ is now only
of order $2n$ in $u$; thus in the relativistic case, two of the
quasiparticle propagating states are projected out.
This is because in the relativistic limit the
spin states naturally coincide with the chiral states.
This is confirmed by noting
that in Eq. (\ref{j22}), for example ($\omega >0$ case), as 
$D^{21}\rightarrow 0$ as $u\rightarrow \omega$, half of the poles
contributing to $J^{22}$ come from (det $D^{11}) \rightarrow 0$.
But for $u\rightarrow \omega$, $D^{11} \rightarrow \omega^2 - u^2
-M^2$. Thus these poles correspond to the vacuum masses; these are
the right handed particle states and left handed antiparticle states
whose masses are unaffected by the left handed effective potential.
As $D^{21}$ reaches zero, the contribution of these poles vanishes completely. 

For $|\omega||\vx-\vy|\gg 1$, Eq. (\ref{glr}) becomes 
\begin{equation}
G_{LR}(\omega,\vx,\vy) \simeq \omega (1 - \mbox{\boldmath $\sigma$}
\cdot \lhat) 
	J(\omega,\vx,\vy). \label{glr2}
\end{equation}
Since we have chosen $\lhat =(\vy_D-\vx_S)/|\vy_D-\vx_S|$ to coincide
with the third spatial 
dimension, 
the only nonzero component of 
$G_{LR}$ is $G_{LR}^{22}$ which for neutrino oscillations is 
$G_{LR}^{22}(|\omega|;\vy,\vx) = 
2|\omega| J^{22}(|\omega|,\vy,\vx)$,
where the spinor space indices are exhibited and the mass/flavor 
indices are suppressed (note that $J^{21}$ vanishes), 
and for antineutrino oscillations is $G_{LR}^{22}(-|\omega|;\vx,\vy) =
-2 |\omega| J^{22}(-|\omega|,\vx,\vy)$.
Thus we see that the Green's function takes the required form 
of Eqs. (\ref{gform}).

The matrix 
$M^2 + 2q\cdot V$ (or $M^2 - 2 q\cdot V$) of \eqr{d22} is precisely
the effective mass matrix $\tilde M^2$ 
appearing in the usual quantum mechanical
model of neutrino (or antineutrino) oscillations. The effective mass matrix
can be diagonalized by a unitary transformation, $\tilde U \tilde M^2
\tilde U^\dagger=1$. 
Thus for neutrino oscillations,
for example, 
\begin{equation}
(D^{22})^{-1}_{\beta\alpha} = \tilde U_{\beta j} \tilde
U_{\alpha j}^* (\omega^2 - u^2 - \tilde m_j^2+i\epsilon)^{-1},
\end{equation}
where $\tilde U_{\beta j}$ are the elements of $\tilde U$ 
and $\tilde m_j^2$ are
the eigenvalues of $\tilde M^2$.
Then Eq. (\ref{glr2}) becomes, using Eq. (\ref{jxy2}),
\begin{equation}
G_{LR}^{\beta\alpha}
(\omega,\vy,\vx) = -|\omega|(1 - \mbox{\boldmath $\sigma$} 
	\cdot \hat{\bf L}) 
	{e^{i|\omega|\hat{\bf L}\cdot(\vy-\vx)} \over 4\pi |\vy_D-\vx_S|} 
	  \sum_j \tilde U_{\beta j} \tilde
U_{\alpha j}^*
e^{-i{\tilde m_j^2\over2|\omega|}|\vy_D-\vx_S|} \label{glr3},
\end{equation}
where we have assumed that the various conditions discussed 
in Sec. II are satisfied. Comparison
of Eq. (\ref{glr3}) with Eqs. (\ref{gform},\ref{oscprob})
shows that the oscillation probability derived here is precisely
the same as that found in the usual quantum mechanical model. 
The antineutrino case works out in a similar manner.

\section{Nonuniform Background}
\label{sec:vback}

In this section we consider the case in which the 
effective potential $V^\mu=V^\mu(\vx)$, that is, we allow
it to vary in space (but not time). 
From Eqs. (\ref{glr3}),(\ref{gform}), and (\ref{oscprob}),
it is clear that in the constant potential case the
portion of the Green's function comprising the
oscillation amplitude obeys a Schr\"{o}dinger-type equation,
the same one used in the standard quantum mechanical picture.
While one might think to simply replace the constant potential
in this Schr\"odinger equation with a spatially
varying one---thus arriving immediately at the standard 
result---we shall go back a little further in order to 
see what is being left out in the process.

In allowing for spatial variation in $V^\mu$, 
Eqs. (\ref{glr}-\ref{jxy})
are unchanged; but Eq. (\ref{jmom}) becomes an integral 
equation, as $J(\omega,\vp)$ must be convolved with the
momentum space dependence of $V^\mu$. 
Since such equations are difficult to deal with 
nonperturbatively, in this section, we 
take a different route of working with a partial
differential equation in coordinate space.  The coordinate space
version of Eq. (\ref{jxy}) is
\begin{equation}
\left[\omega^2 +\nabla^2
 - M^2 - \omega V^0(\vx) -i{\bf V}(\vx)\cdot\right.${\boldmath $\nabla$}$  
 -i ${\boldmath $\sigma$}$ \cdot \left(V^0(\vx)\right.${\boldmath $\nabla$}$
 -i\omega{\bf V}(\vx) + i {\bf V}(\vx) \times
	${\boldmath $\nabla$}$\left.\left.\right)\right] J(\omega,\vx,\vy) = 
	\delta^3(\vx-\vy). \label{jxy3}
\end{equation}  
Unlike in the case of an integral equation, to define the Green's
function using this equation, we must also separately specify the
boundary condition.  We shall assume that the production region is
localized to a region of adiabatically constant potential.
Furthermore, since we expect the virtual particles to all have the
same phase just after being produced, our boundary condition
prescription will be that the the Green's function $J$ asymptote to
the constant potential Green's function on an infinitesimal sphere
centered about $\vy$.\footnote{Note that one must match more than just
the limiting singularity of the Green's function at $\vx=\vy$ to
define a unique solution.}

The form of Eq. (\ref{jxy}), together with our experience in the
vacuum and constant potential cases, suggests that in the
relativistic limit [$M^2/2|\omega|^2 \ll 1,
V|\omega|/|\omega|^2 \ll 
1$ where we suppressed the matrix indices]
we look for solutions of $J$ of the form
\begin{equation}
J(\omega,\vx,\vy)=-{e^{i|\omega||\vx-\vy|}\over 4\pi|\vx-\vy|}
	F(\omega,\vx,\vy). \label{jf}
\end{equation}
With this substitution,
\begin{eqnarray}
(\nabla^2+\omega^2) J &=& \delta^3(\vx-\vy) e^{i|\omega||\vx-\vy|}
	F -  {2|\omega|\,e^{i|\omega||\vx-\vy|}\over 4\pi|\vx-\vy|}
	\left[{1\over 2|\omega|} \nabla^2 F + i (\hat\vr \cdot
	 \mbox{\boldmath $\nabla$} F) - {1 
	\over |\omega||\vx-\vy|}(\hat\vr \cdot
	\mbox{\boldmath $\nabla$} F)\right], \label{ddj} \\
\mbox{\boldmath $\nabla$} J &=& 
	-  {2|\omega|\,e^{i|\omega||\vx-\vy|}\over 4\pi|\vx-\vy|}
	\left[ {i\hat\vr \over 2}F  + {1\over 2|\omega|}
	\mbox{\boldmath $\nabla$} F - {\hat\vr 
	\over 2|\omega||\vx-\vy|} F\right], \label{dj}
\end{eqnarray}
where as before $\hat\vr \equiv {(\vx-\vy)/ |\vx-\vy|}$.
Requiring the first term on the right hand side of Eq. (\ref{ddj})
to cancel the delta function in Eq. (\ref{jxy3}) gives a
boundary condition on $F$, namely (restoring flavor indices) 
\begin{equation}
F^{\beta\alpha}(\omega,\vx,\vy)|_{\vx\rightarrow\vy} = \delta^{\beta\alpha}.
\label{bc}
\end{equation}
Aside from this boundary condition, we are interested in 
well-separated $\vx$ and $\vy$
(specifically $|\omega||\vx-\vy|\gg 1$), so that we
may ignore the last term of Eqs. (\ref{ddj}) and (\ref{dj}).
Then Eq. (\ref{jxy3}) becomes
\begin{eqnarray}
i (\hat\vr \cdot \mbox{\boldmath $\nabla$} F) +
	{1\over 2|\omega|} \nabla^2 F - {1\over 2|\omega|}
	\left[M^2 + \omega V^0 - |\omega|(\hat\vr \cdot {\bf V})
	\right.& &\nonumber\\
\left.- \mbox{\boldmath $\sigma$} 
	\cdot \left(V^0|\omega|\hat\vr - \omega{\bf V} + 
	i |\omega|{\bf V} \times
	\hat\vr \right)\right] F + {\cal O}\left({V|\omega|
	\over |\omega|^2}|\mbox{\boldmath $\nabla$} F|\right)&=&0,
	\label{ddf}	
\end{eqnarray} 
where in accordance with the relativistic condition
$V|\omega|/|\omega|^2 \ll 1$,
we will neglect the terms represented by 
${\cal O}(V|\omega||\mbox{\boldmath $\nabla$} F| 
	/ |\omega|^2)$ in comparison with the
first term of Eq. (\ref{ddf}).

One can distinguish three cases: (1) 
$|\mbox{\boldmath $\nabla$} F| \gg \epsilon F$, where
$\epsilon = V^0 + \tilde{M}^2/(2|\omega|)$ and
$\tilde{M}^2$ denotes the largest mass matrix eigenvalue squared;
(2)
$|\mbox{\boldmath $\nabla$} F| \sim \epsilon F$; and
(3) $|\mbox{\boldmath $\nabla$} F| \ll \epsilon F$.
One can 
argue that case (1) is not interesting since all
terms leading to flavor mixing are rendered negligible.  Case (3) is
also not of present interest because one can argue using
Eq. (\ref{ddf}) that it violates our relativistic assumption.  In
case (2),
$|\nabla^2 F/(2|\omega|)|$ can be neglected compared with
$| \mbox{\boldmath $\nabla$} F|$, 
provided that 
$|\mbox{\boldmath $\nabla$} V^0|/\epsilon^2
\lesssim 1$.
Writing $F' \equiv \hat\vr\cdot\mbox{\boldmath $\nabla$} F$,
Eq. (\ref{ddf}) becomes
\begin{equation}
iF' + {1\over 2|\omega|} D(|\omega|,\vx)F = 0, \label{fprime}
\end{equation} 
where the spinor space elements of $D(|\omega|,\vx)$
 for the particular
case when $\vx$ lies along the third spatial axis $\lhat$ are given by
Eq. (\ref{dmatrix}) with $V\rightarrow V(\vx)$ and $u\rightarrow
|\omega|$.

Since Eq. (\ref{glr2}) holds under the
assumptions of case (2), together with the spatial localization
of the source and detector, we see that 
$G_{LR}^{22}$ is the only nonvanishing component.  Hence, from
Eqs. (\ref{fprime}),(\ref{jf}),(\ref{gform}), and (\ref{oscprob}), we
find that the neutrino oscillation amplitude $H$ obeys the
Schr\"odinger equation
\begin{equation}
iH' = {1\over 2|\omega|}\left[M^2 + 2 q \cdot V(\vx)\right]H,
\end{equation}
where $q = (|\omega|,\hat{\bf L}|\omega|)$. 
Similarly, in the case of antineutrino oscillations
$(\omega < 0)$, the oscillation amplitude obeys
\begin{equation}
i\bar H' = {1\over 2|\omega|}\left[M^2 - 2 q \cdot V(\vx)\right]\bar H.
\end{equation}

Before concluding this section, a remark
regarding the boundary conditions is in order.  Note that $F$ of
\eqr{fprime} satisfying the boundary condition \eqr{bc} is in general
different from $F$ satisfying \eqr{jxy3} with its associated boundary
condition [described just below \eqr{jxy3}].  In particular, although
\eqr{fprime} is valid naively only far away from $\vy$, as we threw
out the last terms of Eqs. (\ref{ddj}) and (\ref{dj}), we still
insisted on the boundary condition \eqr{bc} at $\vy$ to be the same as
the boundary condition that would have been used for the exact
equation.  To justify this, we must show that the terms that we threw
out are negligible even near the origin.  We can argue this by noting
that for 
case (2), 
we are already assuming $|{\boldmath \nabla}
V^0|/\epsilon^2 \lesssim 1$ which turns out to imply (by expanding the
potential to linear order in Taylor series about $\vy$) that in the
relativistic limit the fractional variation of $V^0$ is much smaller
than 1 until $|\omega (\vx - \vy)| \gg 1$ (after which the terms
proportional to $1/|\vx - \vy|$ that we threw out are negligible).
That means that the potential can be treated as a constant until the
terms proportional to $1/|\vx - \vy|$ become negligible.  This implies
that one can place the boundary condition for the varying potential
case on a sphere (centered about $\vy$) on which \eqr{fprime} is valid
using the solution to the constant potential case.  As we saw in the
last section, since the exact solution to the constant potential case
on this sphere (with the appropriate boundary condition) is, up to
relativistically suppressed terms, the same as the solution obtained by
\eqr{fprime} with \eqr{bc}, we can just set the boundary condition for
the varying potential case using \eqr{bc} as well.  Note, however,
that since our argument depends on the adiabaticity of the potential
near the virtual particle production point, in other situations, one
may need to be more cautious with the boundary conditions.

Thus under appropriate conditions the results of the usual simplified
picture are confirmed, including the boundary condition
$H^{\beta\alpha}(\omega,\vx,\vy)|_{\vy\rightarrow\vx}=
\delta^{\beta\alpha}$.

\section{Conclusion}
\label{sec:conc}
Starting from quantum field theory (QFT), we have 
defined a physically meaningful flavor oscillation
probability; determined that portion of the neutrino
propagator that comprises the oscillation amplitude; and
derived the
``Schr\"{o}dinger equation'' for 
that amplitude
in the
presence of spatially varying background matter.  As expected, the
``Schr\"{o}dinger equation'' really corresponds to a time independent
one since its derivation depends on the time-independence of the
effective potential. 
In fact, the usual quantum mechanical approach is only really suited to
problems in which oscillations occur in space only (that is, stationary
systems like that studied here) or time only (e.g. the thermal bath
in the early universe). While we have 
assumed the stationary case here, the basic framework could also be
used to study oscillations in space in the presence of a 
time-dependent background. Ultimately, the description of flavor
oscillating neutrinos in space and time in more general 
systems---i.e. those that do not lend themselves to interpretation 
in terms of a ``source'' and ``detector''---would require a 
formulation in terms of density matrices (cf. Ref. \cite{sigl})
or Wigner functions (cf. Ref. \cite{sireraperez}).  

For situations that can be interpreted in terms of a 
``source'' and ``detector,''
we have also reviewed the conditions under which QFT will be useful in
describing the neutrino oscillation process.\footnote{
Many of these conditions were noted in Refs. 
\cite{rich,grimusstock,camp,kier,cohere,ioapilaf}. }
As long as we are looking in the regime in
which the virtual neutrino goes on shell (rendering any propagator
radiative corrections to be negligible or be merely a constant shift),
the many body aspect of QFT is rendered irrelevant.  In that case,
only the production/detection vertex structure and the spins of the
neutrinos are missing from the usual quantum mechanical treatment.
These, in general, are less difficult to accommodate in the quantum
mechanical treatment than the many body effects.  Still, we believe
them to be more straightforwardly accommodated in the quantum field
theoretical treatment.  For weak interactions, the chiral nature of the
interactions combined with the relativistic nature of the on shell
neutrinos suppresses all but one spin degree of freedom.  Finally, the
smallness of the neutrino mass splittings as well as the neutrinos
going on shell allows one to factor out the production/detection part
of the neutrino scattering process from the ``oscillation'' part,
reducing the problem to the usual 
quantum mechanical 
system
involving a single spinless particle.

To state this another way, we have argued that 
in the context of stationary systems,
the quantum field
theoretic formulation used in this work is not of much use except when
one or more of the following is true: the on-shell neutrino momentum
is nonrelativistic, the production or detection vertices are
non-chiral, the external particle wave packets vary appreciably about
the pole value (value of the wave packet evaluated at the neutrino
momenta) over momentum variations order of the inverse source-detector
distance $1/L$, or the effective mass splittings (determined by the
effective potential including the background matter contributions) are
large compared to the spread in the momenta of the external
production/detection particles.

When the neutrinos are non-relativistic or the interactions are not
chiral, more than one spin contributes per amplitude.  In that case
the usual quantum mechanical treatment must be modified to incorporate
the effects due to the various spin components.  
In particular, in this case 
an oscillation probability cannot be determined
apart from the production and detection processes since 
two neutrino spin contributions are summed before the amplitude is
squared. 
This is, of course,
automatically accounted for in a quantum field theoretic treatment.
Also, if the wave packet varies appreciably about the pole value when the
momentum is varied by ${\cal O}(1/L)$, the wave functions of the
external particles creating and absorbing the virtual neutrino do not
simply factorize out of the oscillation probability amplitude.  Ref.\
\cite{ioapilaf} is a study of one such situation.  They find that the
oscillation amplitude can exhibit a novel ``plane-wave'' behavior.
Just as most effects, this probably can also be accounted for in the
quantum mechanical formulation, but it is much easier in the quantum
field theoretical treatment used in this paper.  The
involvement of the details of the external particles' wave function in
determining the oscillation probability also applies when the
effective mass splitting is much larger than the momentum spread in
the external wave packets, but in such cases, it is not clear whether
any neutrino oscillations can be observed in general because of the
strong suppression of the amplitudes (however, see for example Ref.\
\cite{ioapilaf}).

We here make a few comments regarding Majorana vs. Dirac neutrinos. 
If the neutrinos were Majorana instead of Dirac, the only general
arguments that would change are those dependent upon the existence of a
right handed neutrino.  Although we couched the mathematics in the Dirac
spinor formalism, owing to the assumption of chiral nature of neutrino
interactions and the relativistic limit, none of our general arguments
depended upon the existence of a right handed neutrino.  Hence, 
our conclusions for the relativistic limit 
are also valid for Majorana neutrinos. (Formulae for
``neutrinos'' apply to negative helicity Majorana neutrinos,
while those for ``antineutrinos'' apply to positive helicity 
Majorana neutrinos.)

From our justification of the ``Schr\"{o}dinger equation'' for
stationary, flat spacetime systems, we expect that the heuristic
ansatz used in Ref.\ \cite{cardall} for studying neutrino oscillations
in a stationary curved spacetime to be valid to the extent that the
flat spacetime treatment is valid.  In a nonstationary curved
spacetime, in addition to the time dependence of the potential, there
may arise extra complications of using the S-matrix formalism due to
the nontrivial Bogoliubov transformations of the asymptotic states.
This also deserves further investigation.

\acknowledgements{ CYC thanks Georg Raffelt for helpful
conversations.  We thank the warm hospitality of the
Theoretical Astrophysics group at Fermilab where part of this work was
carried out.  This work was started at the GAAC sponsored meeting held
at the Aspen Center for Physics during the summer of 1998.  CYC is
supported by DOE grant FG02-87ER40317.}

\appendix
\section{Box Quantization and Localization}

In Sec. II we discussed the crucial role that wave packets
play in the localization inherent in neutrino oscillation experiments.
Studies of vacuum oscillations in quantum field theory 
typically begin with some kind of coordinate space description of the
external particles, with the connection
to the underlying momentum space wave packets not explicitly displayed.
Since normalization in quantum field theory is generally fixed
at the level of free-particle momentum eigenstates, fully normalized
event rates cannot be computed unless the connection
to momentum space is made.
The usual practice in the literature has been to ignore this step,
being satisfied with the identification of
the factor called the ``oscillation amplitude'' in the simple
quantum mechanical picture. The complete connection to momentum space
is presumably 
a straightforward exercise, though
perhaps tedious in the context of general wave packets.  However, as
shown in the text, 
the box wave packet allows a
simple, reasonable approximation.  For typical microscopic processes,
a standard trick used to relate the S matrix to rates and cross
sections is to employ the fiction of a box with quantized states,
which reproduces the box wave packet results.  For pedagogical
purposes, we here demonstrate that the use of a similar
fiction---separate boxes confining the external particles at the
source and detector---leads cleanly and directly to an event rate in
the form of Eq. (\ref{rate}).  In this heuristic approach the boxes
serve the double duty of imposing the localization of source and
detector necessary to the existence of oscillations, and the
normalization that gives an absolute event rate.  

For illustration we will 
consider a particular neutrino oscillation process.
The first vertex of the process we consider 
involves the scattering of a charged lepton 
of flavor $\alpha$ off a proton in a source region of volume $V_S$,
producing a final state neutron confined to the source region and 
a virtual neutrino that propagates over all space. At the second vertex,
the virtual neutrino interacts with a
neutron confined to a detection
region of volume $V_D$, producing a proton and charged lepton
of flavor $\beta$ confined to the detection region.    
The free Lagrangian is taken to be 
\begin{eqnarray}
L_0 &=& \sum_s \int_{V_S} d^3x\, \bar\psi_s(i\gamma^\mu\partial_\mu-m)
	 \psi_s +
	\sum_d \int_{V_D} d^3x\,\bar\psi_d(i\gamma^\mu\partial_\mu-m)
	 \psi_d \nonumber \\
& &	+\sum_i \int_{\rm all\ space} d^3x\, \bar\psi_i(i\gamma^\mu\partial_\mu-m)
	 \psi_i, 
\end{eqnarray}
where the index $s$ runs over the particles confined to the source
($\alpha,p,n$), the index $d$ runs over the particles confined to the detector
($\beta,p,n$), and the index $i$ runs over the neutrino mass eigenstates.
The box localization is really a result of complicated classical
potentials preparing the initial state.  The expansions of the source
and detector particle fields are\footnote{Since we are taking
momentum quantization ``seriously'' in this approach, it is convenient
to adopt slightly different normalization conventions from the ones
used in the main text. The differences are easily deduced from 
the explicit expressions for the field expansions and the commutation
relations given above.}
\begin{equation}
\psi_{s,d}(x)={1\over \sqrt{V_{s,d}}}\sum_\sigma\sum_{\vp}
	{1\over \sqrt{2E_{\vp}}}
	\left[a_{\vp,\sigma,(s,d)} u_{\vp,\sigma,(s,d)} e^{-ip\cdot x} 
+ b^\dagger_{\vp,\sigma,(s,d)}v_{\vp,\sigma,(s,d)}e^{ip\cdot x}\right], 
\end{equation}
where the momentum 
sum is over the discrete momenta arising from the application of
periodic boundary conditions to the box, the index $\sigma$ 
is over the spin states, and the commutation relations of the
creation/annihilation operators are of the form 
$[a_{\vp,\sigma,s},a^\dagger_{\vp',\sigma',s'}]=
\delta_{\vp\vp'}\,\delta_{\sigma\sigma'}\,\delta_{s,s'}$. 
In contrast, the free neutrino fields have a continuous momentum spectrum,
\begin{equation}
\psi_i(x)=\sum_\sigma\int {d^3 p \over (2\pi)^{3/2}}{1\over \sqrt{2E_{\vp}}}
	\left[a(\vp,\sigma,i)u(\vp,\sigma,i) e^{-ip\cdot x} 
	+ b^\dagger(\vp,\sigma,i)v(\vp,\sigma,i) e^{ip\cdot x}\right], 
\end{equation}
with $[a(\vp,\sigma,i),a^\dagger(\vp',\sigma',i')]=\delta^3(\vp-\vp')\,
\delta_{\sigma\sigma'}\,\delta_{ii'}$. 

The interaction Lagrangian relevant to the first vertex of the process
described above is
\begin{equation}
L_I= -\lambda 
\int_{V_s} d^3 x\, \bar\psi_\alpha \gamma^\mu(1-\gamma_5) \nu_\alpha
	\bar\psi_p \gamma_\mu(1-g_A\gamma_5) \psi_n +{\rm H.c.},
\end{equation}
where $g\equiv G_F \cos\theta_c /\sqrt{2}$, $G_F$ is the Fermi constant,
$\theta_c$ is the Cabibbo angle, and $g_A\approx 1.26$ is the neutron-proton
axial vector coupling. A similar interaction Lagrangian describes the
interaction in the detector. The relation between the flavor fields and mass
eigenstate fields is $\nu_\alpha = \sum_i U_{\alpha i} \psi_i$, where the
$U_{\alpha i}$ are elements of a unitary matrix.
The amplitude for the neutrino production/detection process is 
\begin{eqnarray}
T_{\alpha\beta} &=& - g^2 \int_{-\infty}^\infty 
	dx^0 \int_{-\infty}^\infty dy^0
	\int_{V_s} d^3 x \int_{V_d} d^3 y 
	{e^{-ip\cdot x}\over\sqrt{V_S}\sqrt{2E_{\vp}}}
	{e^{ip'\cdot x}\over\sqrt{V_S}\sqrt{2E_{\vp'}}}
	{e^{-ik\cdot x}\over\sqrt{V_S}\sqrt{2E_{\vk}}}\nonumber\\
& &	\times {e^{-il\cdot y}\over\sqrt{V_D}\sqrt{2E_{\vl}}}
	{e^{il'\cdot y}\over\sqrt{V_D}\sqrt{2E_{\vl'}}}
	{e^{ik'\cdot y}\over\sqrt{V_D}\sqrt{2E_{\vk'}}}
\left[\bar u(\vl') \gamma^\nu(1-g_A\gamma_5) u(\vl)\right]\nonumber \\
& &\times \left[\bar u(\vk') \gamma_\mu(1-\gamma_5) (i G^{\beta\alpha}(y,x) )
	 \gamma_\nu(1-\gamma_5) u(\vk) \right]
\left[\bar u(\vp') \gamma^\mu(1-g_A\gamma_5) u(\vp)\right].\label{famp1}
\end{eqnarray}
The propagator or Green's function,
$ i G^{\beta\alpha}(y,x) = \langle 
T\{\nu_\beta(y)\bar\nu_\alpha(x)\}\rangle_0$,
with $T\{\}$ and $\langle \rangle_0$ denoting a time-ordered product
and vacuum expectation value respectively, can be expressed in 
momentum space as
\begin{equation}
G^{\beta\alpha}(y,x) = \int {d^4s\over (2\pi)^4}\, 
e^{-is\cdot(y-x)}\,G^{\beta\alpha}(s).
\end{equation}
Integration over $x^0$, 
$y^0$, and $s^0$ in Eq. (\ref{famp1})
 fixes $|s^0|=|E_{\vk}+E_{\vp}-E_{\vp'}|$ and
yields a factor $(2\pi)^2 \delta(E_{\vk}+E_{\vp}+E_{\vl}
-E_{\vk'}-E_{\vp'}-E_{\vl'}) $.

We let $V_S$ and $V_D$ be cubes of sides $L_S$ and $L_D$,
centered on $\vx_S$ and $\vy_D$ respectively, and employ
the approximation of Eq. (\ref{dapprox}).
Because of the V-A lepton currents, the relevant block of the
Green's function is $G_{LR}$. Suppose that this block of the propagator
takes the form of Eq. (\ref{gform}), with $H^{\beta\alpha}$
having only flavor indices.
Changing variables to $\vx'=\vx-\vx_S$
and $\vy'=\vy-\vy_D$, the amplitude in Eq. (\ref{famp1}) can be
expressed
\begin{eqnarray}
T_{\alpha\beta} &=& {i\lambda^2 \,\delta(E_{\vk}+E_{\vp}+E_{\vl}
	-E_{\vk'}-E_{\vp'}-E_{\vl'}) \over 2  (V_s)^{3/2} (V_d)^{3/2}
	|\vy_D-\vx_S| \left(\prod_s \sqrt{2E_s}\right) 
	\left(\prod_d \sqrt{2E_d}\right) } e^{i(\vk+\vp-\vp')\cdot\vx_S}
	e^{i(\vl-\vl'-\vk')\cdot\vy_D} \nonumber \\ 
& &\times H^{\beta\alpha}(\enu,\Yd,\Xs ) 
	\left[\prod_I \Delta(u^I,L_S)\Delta(v^I,L_D)\right] 
	\left[\bar u(\vp') \gamma^\mu(1-g_A\gamma_5) u(\vp)\right]
\nonumber \\
& &\times      \left[\bar u(\vk') \gamma_\mu(1-\gamma_5) u(\vq)^-\right] 
	\left[\bar u^-(\vq) \gamma_\nu(1-\gamma_5) u(\vk)\right]  	   
	 \left[\bar u(\vl') \gamma^\nu(1-g_A\gamma_5) u(\vl)\right],
\label{famp2}
\end{eqnarray} 
where the index $I$ is over the three momentum directions, ${\bf
u}\equiv\vk+\vp-\vp'-\vq$, ${\bf
v}\equiv\vl-\vl'-\vk'+\vq$, $\Delta(w,a)\equiv (2/w)\sin
(wa/2)$, and $\vq = \enu\,\hat{\bf L}$.

An event rate is obtained by squaring the amplitude; interpreting
one of the energy delta functions as $T/(2\pi)$, where $T$ is a
time interval; and dividing by $T$. At this stage we will also
make the approximation of a continuum of states in the source and
detector, i.e. take $(L_S/2),(L_D/2)$ to be large. In this limit
$\Delta(w,a)\rightarrow (2\pi)\delta(w)$, and 
$[\Delta(w,a)]^2\rightarrow [(2/w)\sin (wa/2)|_{w\rightarrow 0}]
(2\pi)\delta(w) = 2\pi a\, \delta(w)$. Finally, in the continuum
limit there is a phase space factor of the form $d^3p\, V/(2\pi)^3$
for each final state particle. The event rate obtained from Eq. 
(\ref{famp2}) is then
\begin{eqnarray}
d\Gamma_{\alpha\beta} &=&{\lambda^4 
	\,\delta(E_{\vk}+E_{\vp} +E_{\vl}
	-E_{\vk'}- E_{\vp'} -E_{\vl'})   
	\over 4 (2\pi)^4 V_s\, L^2
	\left(\prod_s 2E_s\right) \left(\prod_d 2E_d\right)}
	\left|H^{\beta\alpha}(\enu,\Yd,\Xs )\right|^2 \nonumber\\
& &\times	\delta^3(\vp+\vk-\vp'-\vq)
	\delta^3(\vl+\vq-\vl'-\vk') \,d^3p' \, d^3l' \, d^3 k'
 \nonumber \\
& &\times
\sum_{\rm spins}\left[\left|\bar u(\vp') \gamma^\mu(1-g_A\gamma_5) u(\vp)
         \bar u(\vk') \gamma_\mu(1-\gamma_5) u(\vq)\right|^2\right.
\nonumber\\ 
& &\times\left.	\left|\bar u(\vq) \gamma_\nu(1-\gamma_5) u(\vk)  	   
	   \bar u(\vl') \gamma^\nu(1-g_A\gamma_5) u(\vl)\right|^2\right].
	\label{rate3}    
\end{eqnarray}
Using standard plane wave methods to compute the neutrino production rate and 
cross section factors in Eq. (\ref{rate}), it is easy to verify that
Eq. (\ref{rate3}) is equivalent to Eq. (\ref{rate}), with 
$P_{\nuanub} = \left|H^{\beta\alpha}(\enu,\Yd,\Xs )\right|^2$. As
noted in the text, explicit computations 
yield the
oscillation probabilities found in the usual quantum mechanical model.

Eq. (\ref{rate3}) represents the rate per source proton, source
charged lepton, and detector proton. To get the total experimental
rate it is necessary to sum over these source and detector particles.
Employing the usual classical distribution function, for uniform spatial
distribution the number of particles of type $x$ is
\begin{equation}
N_x = V \int {d^3p\over (2\pi)^3} f_x(\vp),
\end{equation}
so that the total event rate is
\begin{equation}
d\Gamma_{\rm exp}=\int V_S {d^3p\over (2\pi)^3}\, f_p(\vp)\,
	V_S {d^3k\over (2\pi)^3}\, f_\alpha(\vk)\,
	V_D{d^3l\over (2\pi)^3}\, f_p(\vl)\,
	d\Gamma_{\alpha\beta}. 
\end{equation}
We note that one of the factors of $V_S$ cancels, leaving one factor
each of $V_S$ and $V_D$. These remaining
two volume factors are replaced by $d^3{\vx}_S$ and $d^3{\vy}_D$, 
and integrations
over these source and detector positions are performed. While we have 
explicitly shown how this works out for our particular chosen process,
for any neutrino production/detection process one factor of $V_S$
and $V_D$ will remain at the end.



\begin{thebibliography}{99}
\frenchspacing
\def\prpts#1#2#3{Phys. Reports {\bf #1}, #2 (#3)}
\def\prl#1#2#3{Phys. Rev. Lett. {\bf #1}, #2 (#3)}
\def\prd#1#2#3{Phys. Rev. D {\bf #1}, #2 (#3)}
\def\prc#1#2#3{Phys. Rev. C {\bf #1}, #2 (#3)}
\def\plb#1#2#3{Phys. Lett. {\bf #1B}, #2 (#3)}
\def\npb#1#2#3{Nucl. Phys. {\bf B#1}, #2 (#3)}
\def\apj#1#2#3{Astrophys. J. {\bf #1}, #2 (#3)}
\def\apjl#1#2#3{Astrophys. J. Lett. {\bf #1}, #2 (#3)}

\bibitem{sk} The Super-Kamiokande Collaboration: Y.Fukuda et al.,
	\plb{433}{9}{1998}; The Super-Kamiokande 
	Collaboration: Y.Fukuda et al.,
	\plb{436}{33}{1998}; The 
	Super-Kamiokande Collaboration: Y. Fukuda et al., 
	\prl{81}{1158}{1998}; Erratum \prl{81}{4279}{1998}.

\bibitem{solrev} Recent reviews and updates
	include e.g. Y. Suzuki, Space Sci. Rev.
	{\bf 85}, 91 (1998); V. Barger and K. Whisnant,
	hep-ph/9812273;
	J. N. Bahcall, P. I. Krastev, and
	A. Y. Smirnov, \prd{58}{096016}{1998}; 
	G. L. Fogli, E. Lisi, and D. Montanino,
	Astropart. Phys. {\bf 9}, 119 (1998);
	N. Hata and P. Langacker, 
	\prd{56}{6107}{1997}.

\bibitem{atmrev} Recent updates include e.g.
	 G.L. Fogli, E. Lisi, A. Marrone, and G. Scioscia,
	\prd{59}{033001}{1999};
	T. Teshima and T. Sakai, Prog. Theor.
	Phys. {\bf 101}, 147 (1999); 
	O. Yasuda, \prd{58}{1301}{1998};
	V. Barger, T. J. Weiler,
	and K. Whisnant, \plb{440}{1}{1998}; Y. Oyama,
	\prd{57}{R6594}{1998}.

\bibitem{skatmint} The Super-Kamiokande Collaboration: Y. Fukuda et al., 
	\prl{81}{1562}{1998}. 

\bibitem{longbl} For overviews see e.g. G. Barenboim
	and F. Scheck, hep-ph/9812351; K. Zuber, presented at
	``International Workshop on Simulation and Analysis Methods for 
	Large Neutrino Detectors,'' hep-ex/9810022; G. Battistoni
	and P. Lipari, presented at ``Vulcano workshop on Frontier
     Objects in Astrophysics and Particle Physics,'' hep-ph/9807475;
	S.M. Bilenky, C. Giunti, and W. Grimus, presented at
	``XXXIIInd Rencontres de Moriond: Electroweak
     Interactions and Unified Theories,'' hep-ph/9805387; and
	G.L. Fogli and E. Lisi, \prd{54}{3667}{1996}.

	
\bibitem{solsig} SNO  
	(e.g. A. McDonald for the SNO Collaboration,
	talk presented at ``Neutrino '98, XVIII International Conference
	on Neutrino Physics and Astrophysics,'' to appear in the
	proceedings), a heavy water detector, is expected to have 
	a measurable neutral current event rate if the solar neutrino
	problem is solved by conversion of  
	$\nu_e$ to $\nu_\mu$ or $\nu_\tau$.
	Both Super-Kamiokande and SNO
	will look for deviations from the expected
	recoil electron spectrum, as well as anomalous seasonal
	or diurnal time variations. The status of the 
	Super-Kamiokande data on
	these signals is reported in The Super-Kamiokande Collaboration,
	hep-ex/9812011 and hep-ex/9812009.

\bibitem{lsnd}LSND Collaboration: C. Athanassopoulos et al., 
	\prl{81}{1774}{1998}; LSND Collaboration: 
	C. Athanassopoulos et al., nucl-ex/9706006.

\bibitem{wolf} L. Wolfenstein, \prd{17}{2369}{1978}.

\bibitem{ms} S. Mikheyev and S. Yu. Smirnov, Nuovo Cimento
	Soc. Ital. Fis. C {\bf 9}, 17 (1986).

\bibitem{rich} J. Rich, \prd{48}{4318}{1993}.

\bibitem{grimusstock} W. Grimus and P. Stockinger, \prd{54}{3414}{1996}.

\bibitem{camp} J. E. Campagne, \plb{400}{135}{1997}.

\bibitem{kier} K. Kiers and N. Weiss, \prd{57}{4418}{1998}.

\bibitem{cohere} W. Grimus and P. Stockinger, \prd{59}{013011}{1999}.

\bibitem{ioapilaf} A. Ioannisian and A. Pilaftsis, \prd{59}{053003}{1999}.

\bibitem{sireraperez} M. Sirera and A. P\'{e}rez, hep-ph/9810347.

\bibitem{peskin}M. E. Peskin and  D. V. Schroeder,
	{\em Introduction to Quantum Field Theory} 
	(Reading, Addison-Wesley, 1995). 

\bibitem{kuorev1} T. K. Kuo and J. Pantaleone, Rev. Mod. Phys. {\bf
61}, 937 (1989).

\bibitem{notraf} D. N\"{o}tzold and G. Raffelt, \npb{307}{924}{1988}.

\bibitem{fuller}G. M. Fuller, R. W. Mayle, J. R. Wilson, and D. N. 
	Schramm, \apj{322}{795}{1987};
	G. M. Fuller, R. W. Mayle, B. S. Meyer, and J. R. Wilson,
	\apj{389}{517}{1992}.

\bibitem{qf1}Y.-Z. Qian, G. M. Fuller, G. J. Mathews, R. W. Mayle,
	J. R. Wilson, and S. E. Woosley, \prl{71}{1965}{1993}.

\bibitem{qf2}Y.-Z. Qian and G. M. Fuller, \prd{51}{1479}{1995}.

\bibitem{sigl}G. Sigl and G. Raffelt, \npb{406}{423}{1993}.

\bibitem{cardall} C. Y. Cardall and G. M. Fuller, \prd{55}{7960}{1997}.


\end{thebibliography}
\end{document}